\documentclass[fleqn,usenatbib]{mnras}

\usepackage{amssymb,amsmath}

\newcommand{\flux}{\,erg\,s$^{-1}$\,cm$^{-2}$} %
\newcommand{\lumi}{\,erg\,s$^{-1}$} %

\usepackage[T1]{fontenc}
\usepackage{ae,aecompl}

\usepackage{graphicx}	
\usepackage{amsmath}	
\usepackage{amssymb}	

\usepackage[symbol]{footmisc}
\usepackage{fp}
\usepackage{multirow}
\usepackage{footnote}
\usepackage{tabularx}
\usepackage{lscape}

\title[Coronae and exoplanet helium transits]{Helium absorption in exoplanet atmospheres is connected to stellar coronal abundances
}

\author[K. Poppenhaeger]{
K. Poppenhaeger$^{1, 2}$\thanks{E-mail: kpoppenhaeger@aip.de}
\\
$^{1}$Leibniz Institute for Astrophysics Potsdam (AIP), An der Sternwarte 16, 14482 Potsdam, Germany\\
$^{2}$Universit\"at Potsdam,  Institut f\"ur Physik und Astronomie,  Karl-Liebknecht-Stra{\ss}e 24/25, 14476 Potsdam, Germany}

\date{Accepted 2022 February 17. Received 2022 February 17; in original form 2021 September 14}

\pubyear{2022}

\begin{document}
\label{firstpage}
\pagerange{\pageref{firstpage}--\pageref{lastpage}}
\maketitle

\begin{abstract}
Transit observations in the helium triplet around 10830 Angstrom are a successful tool to study exoplanetary atmospheres and their mass loss. Forming those lines requires ionisation and recombination of helium in the exoplanetary atmosphere. This ionisation is caused by stellar photons at extreme ultra-violet (EUV) wavelengths; however, no currently active telescopes can observe this part of the stellar spectrum. The relevant part of the stellar EUV spectrum consists of individual emission lines, many of them being formed by iron at coronal temperatures. The stellar iron abundance in the corona is often observed to be depleted for high-activity low-mass stars due to the first ionisation potential (FIP) effect. I show that stars with high versus low coronal iron abundances follow different scaling laws that tie together their X-ray emission and the narrow-band EUV flux that causes helium ionisation. I also show that the stellar iron to oxygen abundance ratio in the corona can be measured reasonably well from X-ray CCD spectra, yielding similar results to high-resolution X-ray observations. Taking coronal iron abundance into account, the currently observed large scatter in the relationship of EUV irradiation with exoplanetary helium transit depths can be reduced, improving the target selection criteria for exoplanet transmission spectroscopy. In particular, previously puzzling non-detections of helium for Neptunic exoplanets are now in line with expectations from the revised scaling laws.
\end{abstract}




\section{Introduction}

Atmospheres of exoplanets can be detected and characterised in a variety of ways. While there are some possibilities to observe atmospheric features of non-transiting planets (e.g.\ \citealt{Snellen2014, Birkby2017}), the overwhelming majority of exoplanets whose atmospheres have been observed successfully are exoplanets in transiting orbits around their host stars (e.g.\ \citealt{Charbonneau2002, Tinetti2007, Bean2010, Kreidberg2014, Keles2020}). Of particular interest is the fact that some exoplanets appear to have eva\-po\-ra\-ting atmospheres, which can produce gaseous tails streaming behind the planet \citep{Kulow2014, Ehrenreich2015, Nortmann2018}, or in some cases may even produce bow shocks in front of the planet on its orbit \citep{Llama2011, Cauley2015}. 

One method of observing such extended and escaping atmospheres is through transmission spectroscopy, often by targeting spectral lines of volatile atomic species such as hydrogen and helium. Typical targets are exoplanets with a thick gaseous envelope, such as hot Jupiters and Neptunes. Observations of the hydrogen Ly-$\alpha$ line in the ultraviolet have led to detections of escaping atmospheres for a handful of exoplanets \citep{Vidal-Madjar2003, Lecavelier2010, Kulow2014, Ehrenreich2015, Bourrier2018}. However, such detections are complicated by the fact that hydrogen in present in the universe all along the line of sight of such observations. Specifically, the stellar Ly-$\alpha$ photons undergo absorption by the extended planetary atmosphere, followed by absorption by the interstellar medium in the line of sight between the exoplanet and the observer. Finally, the geocorona contributes to the observed Ly-$\alpha$ line with an emission component, so that the resulting observed line profile needs to be analysed with great care in order to isolate the exoplanetary atmospheric effect \citep{Vidal-Madjar2003}.

\cite{Oklopcic2018} presented calculations showing that the metastable lines of helium in the infrared at 10830\,\AA\ should produce strong observable signatures of exoplanetary absorption during transits, which was already proposed by \citet{Seager2000}. In contrast to the hydrogen Ly-$\alpha$ line, this line of helium is not vulnerable to absorption by the interstellar medium, because the lower level of the line transition is not the ground state, but a metastable level reached after ionisation and recombination. The helium in the interstellar medium is typically not present in this state and therefore does not absorb at 10830\,\AA. Subsequently, several detections of exoplanetary atmospheric absorption in the He\textsc{I}10830 line have been made in the past two years (see for example \citealt{Mansfield2018, Nortmann2018, Allart2019, Alonso-Floriano2019, Guilluy2020, Paragas2021}).

How helium atoms reach the lower state of the metastable He\textsc{I}10830 transition is a matter of detailed research in the solar and stellar community (see \citet{Linsky2017} for a review). The Sun displays some contrast in the He10830 line across the solar disk, with active regions being darker in He\textsc{I}10830 than the quiescent solar atmosphere. Understanding the profiles of the solar helium lines, both the resonance lines in the UV and the metastable lines in the near-infrared, is non-trivial. Early models of the solar transition region struggled to produce the required ionisation fraction to match observed line profiles and needed to impose \textit{ad hoc} constraints on local temperatures \citep{Jordan1975, Vernazza1981}. It turned out that ambipolar diffusion -- where ions diffuse into the cooler, less ionised regions and atoms diffuse into the hotter, more ionised regions of the transition region -- as well as backwarming from the corona play an important role in the ionisation and excitation balances of the transition region \citep{Fontenla1990, Fontenla1993}.

\citet{Andretta1997} argued that in the solar case the metastable state of helium can be reached with two main pathways, one being via photoionisation plus recombination of helium, the other being collisional excitation from the ground state. While the radiative transition from the metastable state to the ground state is slow, electron collisions can still depopulate the metastable state by transferring the atom either directly to the ground state or alternatively to the excited singlet state, which easily decays to the ground state as well.
The photoionisation and recombination pathway requires photons with energies typically found in the stellar corona, and lines that are formed via this pathway display spatial features on the Sun that are co-located with coronal features. Collisional excitation requires a high enough density of electrons with sufficient energy, and lines formed via this pathway tend to display features co-located with solar chromospheric structures. For the Sun,  features visible in the He\textsc{I}10830 line are observed in spatial agreement with both chromospheric and coronal structures \citep{Dupree1996, Libbrecht2019}, indicating that both pathways take place.

For stellar samples \citet{Sanz-Forcada2008} argued that photoionisation-recombination is the dominant process for giant stars, while main-sequence cool stars with their dense chromospheres can populate the metastable level through a mix of both pathways. However, there is observational evidence that stars with a higher activity level also display deeper He\textsc{I}10830 lines, indicating that the formation of metastable helium in stellar atmospheres is still dominated by the photoionization-recombination pathway, where stellar EUV photons ionize the helium first. This was concluded from observations that show that the helium triplet absorption is well correlated with stellar high-energy emission for F, G and K stars \citep{Zarro1986}. Recent studies of M dwarfs and their He\textsc{I}10830 triplet demonstrate that the He\textsc{I}10830 lines become deeper during flares of the stars, further strengthening the assumption that an EUV-driven photoionization plus recombination is a dominant pathway of the formation of this line in stars.

In contrast, the atmospheres of exoplanets display very different temperatures and densities. \citet{Oklopcic2018} calculates that collisional excitation is expected at a much lower rate than the EUV-driven photoionisation and recombination in a planetary atmosphere. \citet{Oklopcic2019} states that stars with higher magnetic activity, i.e.\ higher X-ray and UV (together, XUV) emission levels, should produce more helium in the relevant metastable state in an exoplanetary atmosphere. This trend was observationally confirmed for a small sample of exoplanets by \citet{Nortmann2018}, who found larger effective exoplanet radii at the He\textsc{I}10830 wavelength compared to the broadband optical ones with higher XUV irradiation of those exoplanets. Later works, such as \citet{Kasper2020}, reported the same trend for larger samples of exoplanets.

\begin{figure}
\includegraphics[width=\columnwidth]{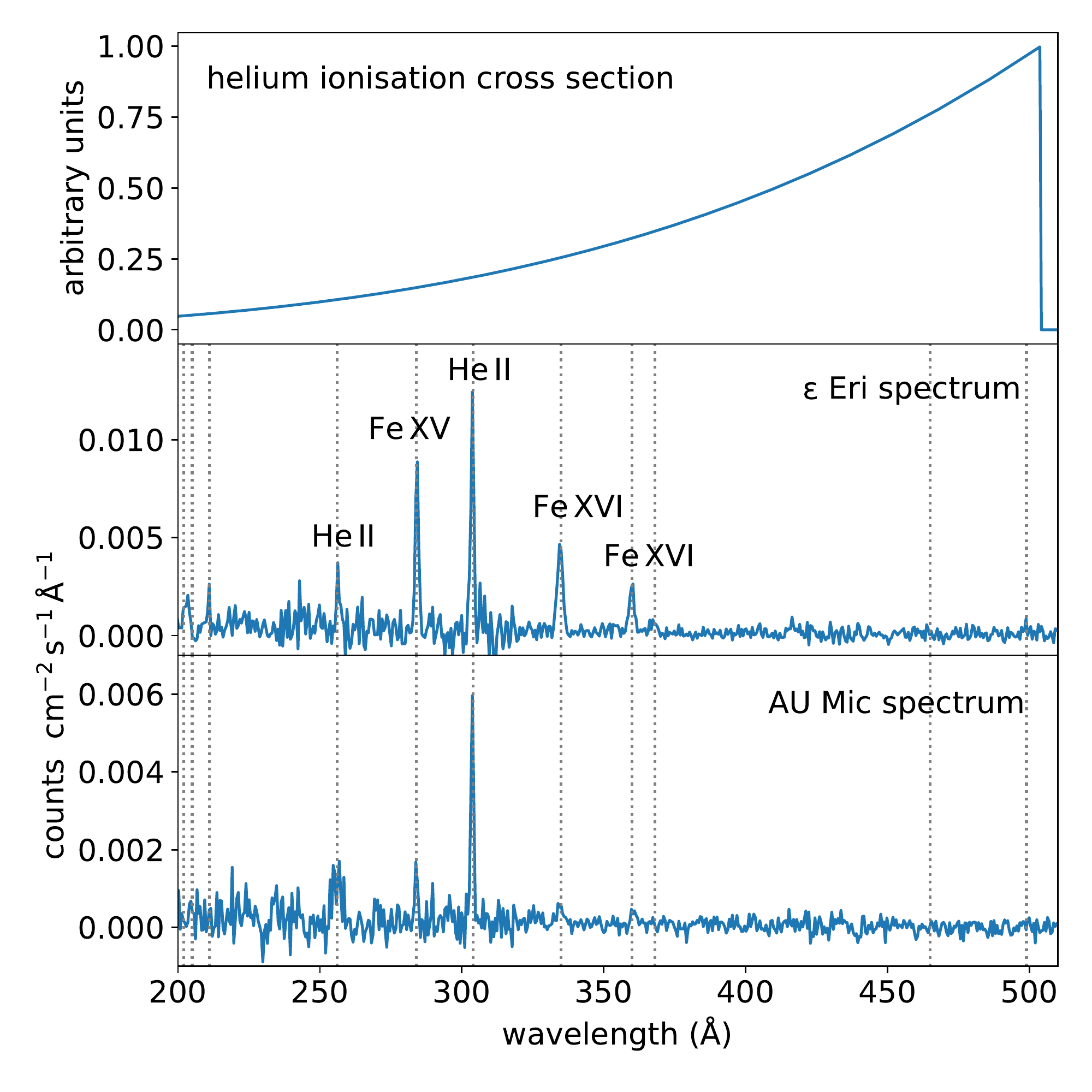}
\caption{Top panel: the relative photoionisation cross section of helium in its ground state. Middle panel: EUVE spectrum of $\epsilon$ Eri, with the helium line at 304\,\AA\ and several strong iron lines being visible. Bottom panel: EUVE spectrum of AU Mic, where the iron lines are much weaker in comparison to the He\textsc{II} lines. The wavelengths of the typically strongest emission lines as given in Table~\ref{tab:lines} are marked by vertical dotted lines.}
\label{fig:spectra_cross}
\end{figure}

The stellar spectrum has a strong effect on the formation of the exoplanetary He\textsc{I}10830 absorption line. The photoionisation threshold for helium in the ground state of the exoplanetary atmosphere is at 504\,\AA\ (Fig.~\ref{fig:spectra_cross}, top panel). The photoionisation cross section of photons with shorter wavelengths decreases by a factor of about 20 for wavelengths of about 200\,\AA\ and below \citep{Yan1998}. The photoionisation of exoplanetary helium by stellar photons therefore happens practically entirely in the extreme-ultraviolet (EUV) regime between 504 and 200\,\AA. While scaling relations between the X-ray and broad-band EUV output of stars have been investigated \citep{Sanz-Forcada2011, King2021}, the narrow EUV wavelength range relevant for He\textsc{I}10830 absorption has not been studied for its relation to X-ray luminosity yet. The observed correlation of He\textsc{I}10830 absorption with X-ray irradiation is expected because EUV and X-ray emission are driven by the same phenomenon in cool stars, namely magnetic activity. The stellar X-ray photons themselves, however, do not contribute directly to the production of exoplanetary helium in the correct state for He\textsc{I}10830 absorption. Since there are currently no astrophysical observatories operating in the wavelength regime between 200 and 504\,\AA, investigating scaling relationships between stellar soft X-rays and the relevant narrow-band EUV range is therefore important.

In this work I aim to characterize the relationship bet\-ween helium ionisation in exoplanetary atmospheres caused by the narrow-band EUV emission of stars -- as a precursor step to the recombination process which yields helium in the ground state of the He\textsc{I}10830 line -- , and the stellar X-ray emission. Present-day X-ray observatories observe sources routinely at energies of 0.2~keV and above, corresponding to wavelengths of about 60\,\AA\ and below. The spectra of cool stars in the EUV and X-rays are predominantly line emission spectra from bound-bound transitions, and I will show that many of those lines originate from the same coronal plasma. Particularly, emission lines of iron will be shown to play an important role. The production rate of metastable helium in exoplanetary atmospheres can be linked not only the to ionising EUV irradiation, but also to the more easily measurable soft X-ray irradiation, and specifically the temperature and iron abundance of the coronal plasma.

A secondary goal of this work is to explore the potential of medium-resolution CCD spectra in X-rays to determine relative iron abundances in coronal plasmas. Traditionally, high-resolution X-ray spectra are used to fit abundances of a large number of elements. However, a few elements that produce several strong lines in soft X-rays, such as iron, oxygen, and neon, can have measurable effects also in lower resolution spectra. This can open up a range of only moderately X-ray bright stars to coronal abundance determinations which would be hard to obtain otherwise.

I describe the used observations and the respective data analysis procedures in sections~\ref{obs}, the results are given in section~\ref{res}, and a discussion in the context of exoplanetary atmosphere studies is given in section~\ref{disc}.

\section{Observations and Data Analysis}\label{obs}

I chose to analyse a sample of stars from the Extreme Ultra-Violet Explorer (EUVE) stellar atlas \citep{Craig1997}, specifically the main-sequence stars from spectral type A to M, plus the only pre-main sequence cool star contained in that atlas, AB Dor. The sample is listed in Table~\ref{tab:sample}. I analysed their EUV emission as well as their X-ray emission.

The EUVE stellar atlas contains several stars that are close binaries, such as AB Dor and BF Lyn. Those binary stars are neither spatially resolved in their EUVE observations nor in the analysed X-ray observations. Therefore I did not attempt to distribute the observed emission at EUV and X-ray wavelengths between their stellar components, but used the emission from the binary system as a whole in the following analysis.

\subsection{Extreme ultraviolet data}

The Extreme Ultra-Violet Explorer (EUVE) was a space telescope operated by NASA from 1992 to 2001 \citep{EUVE}. It remains the most up-to-date high-resolution spectrometer in the extreme-UV. EUVE carried three slitless imaging spectrometers, covering overlapping spectral regions. Specifically, the short-wavelength (SW) spectrometer covered a range of 90--170\,{\AA}, and the medium (MW) and long (LW) spectrometers covered the 140--380\,{\AA} and 280--760\,{\AA} range, respectively. The spectral resolution was ca.\ 0.5\,{\AA}, 1\,{\AA}, and 2\,{\AA} for the SW, MW, and LW spectrometers, respectively. To remove spectral signatures of the Earth's airglow and geocoronal lines, EUVE used filters for all three spectrometers and wire-grid collimators for the MW and SL spectrometers.

The archival EUVE spectra of the stellar atlas stars are available in their reduced form via the Mikulski Archive for Space Telescopes (MAST)\footnote{\url{https://archive.stsci.edu/prepds/atlaseuve/datalist.html}}. These spectra are co-added for targets where several observations were conducted.
The archive provides spectra in units of photons\,cm$^{-2}$\,s$^{-1}$\,\AA$^{-1}$, in a single file each for each of the three arms of the spectrometer, and already corrected with respect to the effective area of EUVE. The relevant wavelength region between 200 and 504\,\AA\ is covered by the MW and LW spectrometer arms. Since the noise level in each arms increases close to the ends of the covered wavelength range, I chose to stitch together the MW and LW data at a wavelength of 330\,\AA, meaning that data points with shorter wavelengths come from the MW arm and data points with longer wavelength come form the LW arm. For plotting purposes only, the MW data points were binned by a factor of two to have the same bin size along the x-axis in the plots.

Two examples of the spectra in the 200--500\,\AA\ region are shown in Fig.~\ref{fig:spectra_cross}, middle and lower panel. The spectrum of the high-activity M dwarf binary EQ Peg is dominated by the helium emission lines at 256 and 304\,\AA, while the spectrum of the moderately active K dwarf $\epsilon$~Eri also shows other emission lines, mainly stemming from iron. It needs to be pointed out that the helium ionisation cross section, which is displayed in the top panel of the same figure, also determines how much of the intrinsic stellar spectrum gets absorbed by the interstellar medium (ISM) on its way to the observer. Therefore, stellar emission lines closer to the helium ionisation edge will already be strongly suppressed in the observed spectra. I discuss the uncertainties this introduces into the analysis in section~\ref{ism_abs_quant}.

\begin{figure}
\includegraphics[width=0.48\textwidth]{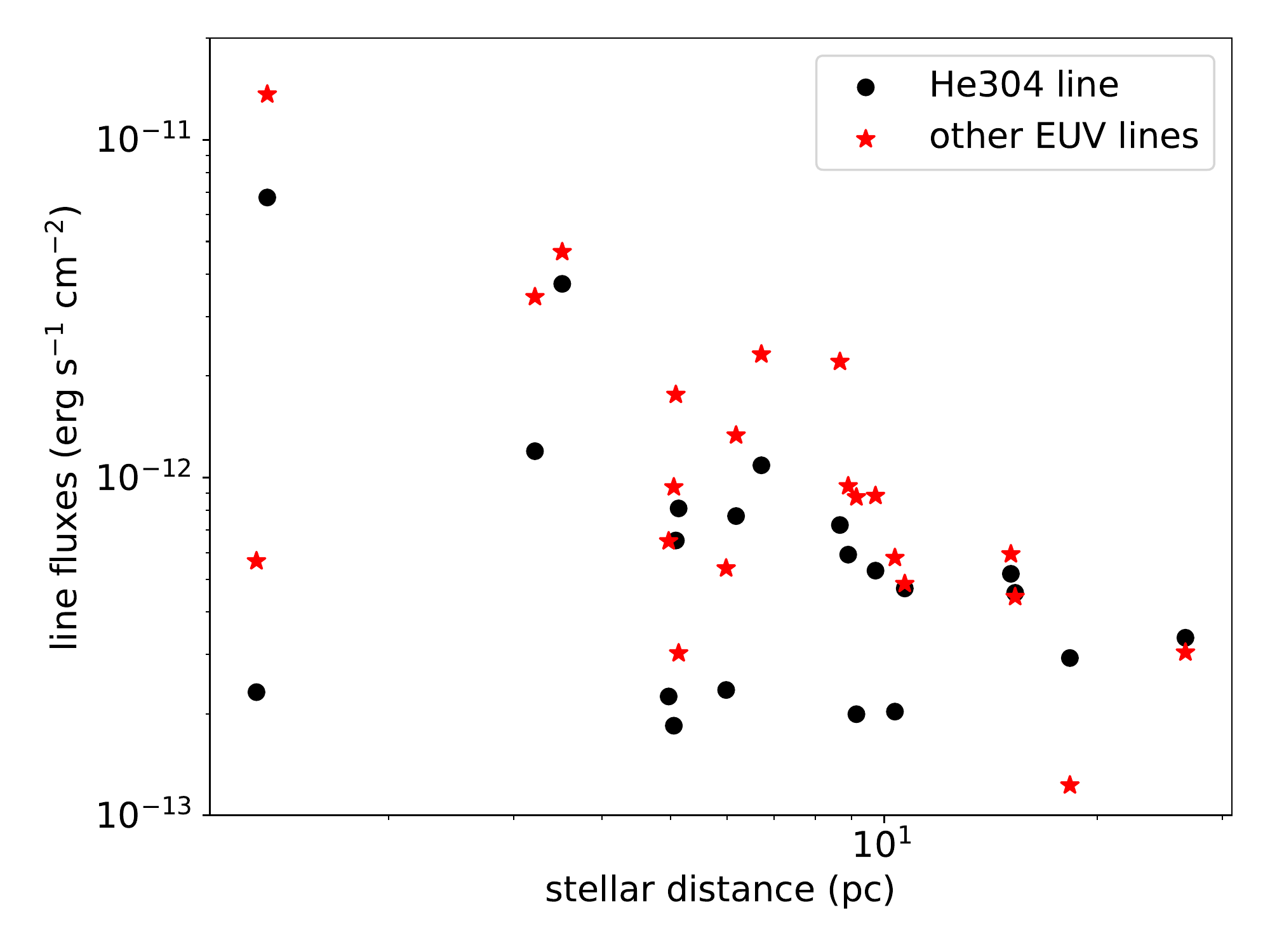}
\includegraphics[width=0.48\textwidth]{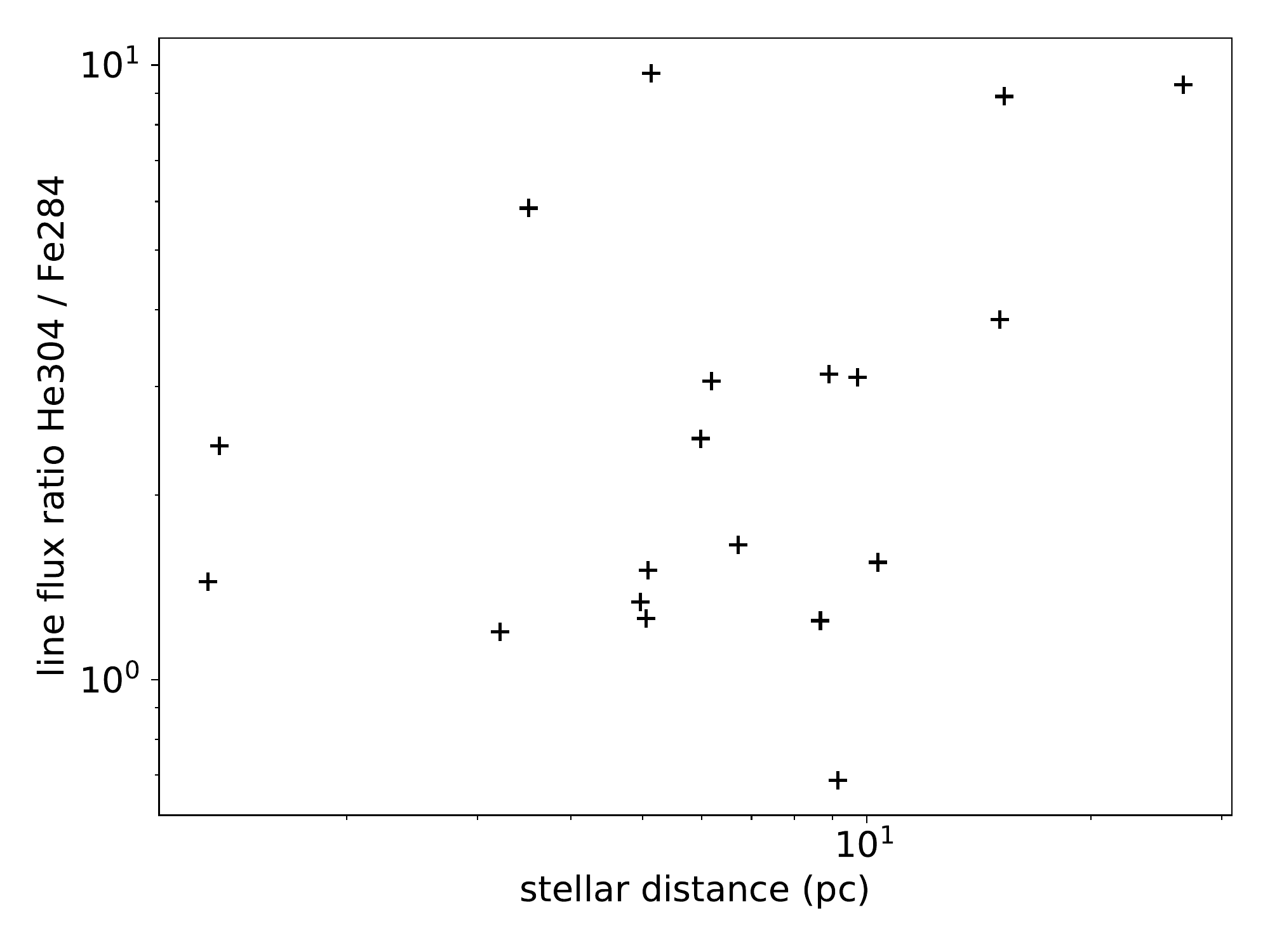}
\includegraphics[width=0.48\textwidth]{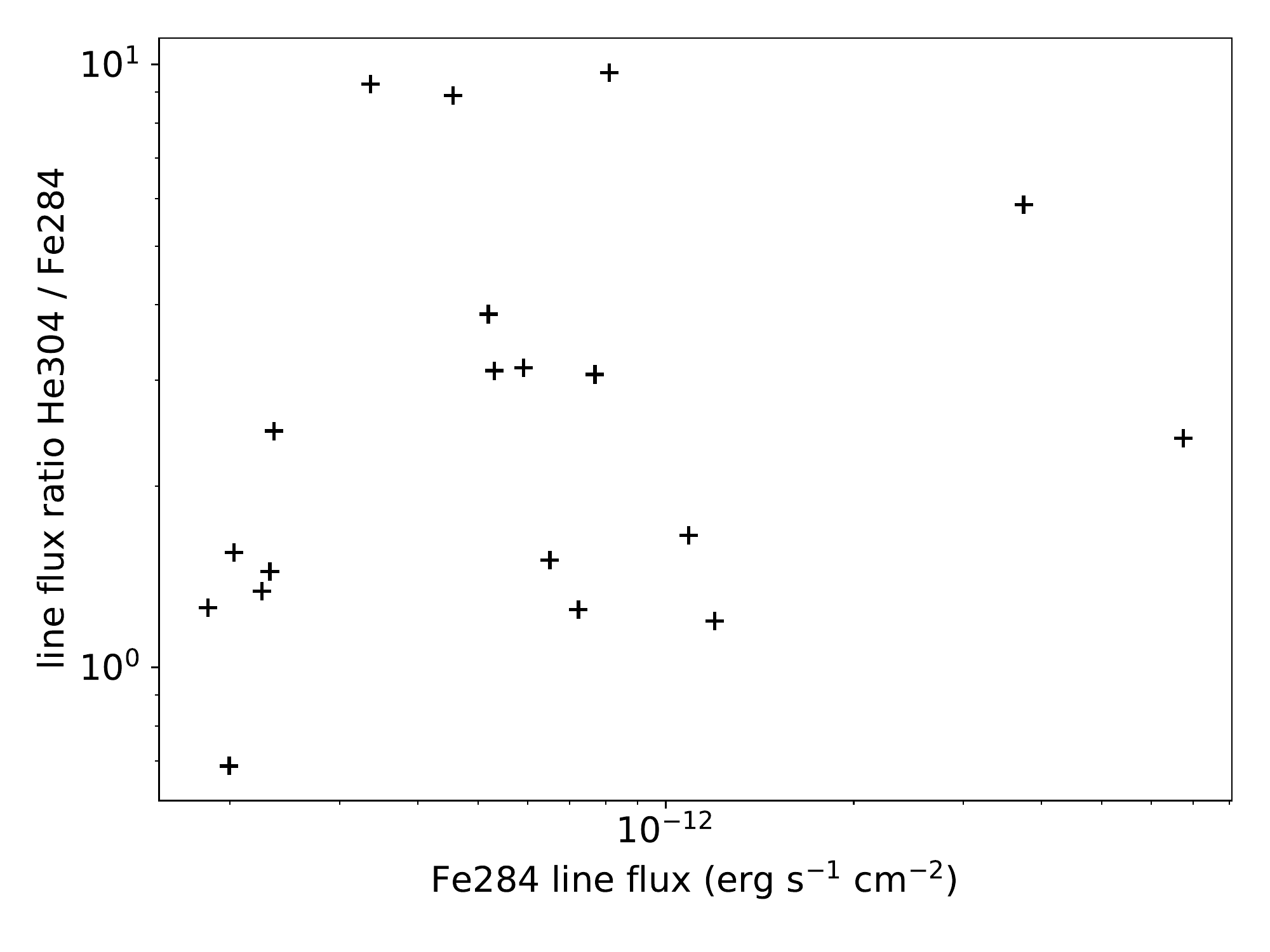}
\caption{\textit{Top:} EUV line fluxes in the He\textsc{II}304 line and other EUV lines listed in Table~\ref{tab:lines} versus stellar distance. If geocoronal contamination were a dominant effect in the observed He\textsc{II}304 lines, the trend of He\textsc{II}304 with distance would differ strongly from the trend seen for the other line fluxes, which is not the case. \textit{Middle:} The line ratio of He\textsc{II}304 to the nearby Fe\textsc{XV}284 line as a function of distance. \textit{Bottom}: The same line ratio as a function of the Fe\textsc{XV}284 line flux. If the He\textsc{II}304 line was mainly geocoronal, we should see an increase of the line ratio with distance as geocoronal emission does not get suppressed by stellar distance, and a decrease with increasing Fe\textsc{XV}284 flux. However, this is not the case, demonstrating that the He\textsc{II}304 emission in the sample is primarily stellar.}
\label{fig:lines_vs_dist}
\end{figure}

The used spectral atlas data sets are already cleaned from high-background time stretches \citep{Craig1997}. One potential concern is the He\textsc{II} line at 304\,\AA, which is contained in the spectral range we are interested in. This line can contain contamination from Earth's geocorona, which to a certain extent should already be limited by the exclusion of high background times in the reduced spectra. To check that geocoronal emission is not a dominant effect in the spectra, I plotted photon line fluxes versus stellar distance, for the He\textsc{II} line as well as for the summed-up fluxes of other EUV emission lines that do not have a potential geocoronal contamination (see Table~\ref{tab:lines}). If geocoronal emission were a dominant contribution, we would expect the He\textsc{II} line flux not to follow the same trend as the other line fluxes with distance, especially at larger distances where the intrinsic stellar line flux would be expected to be weaker compared to a hypothetical contamination. However, Fig.~\ref{fig:lines_vs_dist} (top panel) clearly shows that the He\textsc{II} line follows the same trend as the other lines, indicating that geocoronal contamination of the line is not a dominant factor in the used spectra. As an additional test I calculated the line ratio of the He304 line versus the Fe284 line, as shown in the middle and bottom panels of Fig.~\ref{fig:lines_vs_dist}. If geocoronal flux in the He304 were a dominant contribution in the observed spectra, we would expect to see an increase of this ratio with stellar distance, because the geocoronal He304 emission would not be affected by the distance to the target. Likewise, we would expect to see a decrease of this ratio with increasing Fe284 flux, because the nearer and therefore brighter stars would experience a smaller relative contamination by geocoronal He304 emission. However, none of those trends are evident in Fig.~\ref{fig:lines_vs_dist}, and I conclude that geocoronal contamination is not a strong factor in the spectra analysed here.

\subsection{X-ray data}

The X-ray observations analysed here were collected with the XMM-Newton space telescope \citep{Jansen2001}, with details being given in the following section. XMM-Newton carries three types of instruments onboard: a set of three X-ray CCD cameras \citep{EPIC_mos, EPIC_pn}, collectively called EPIC, which provide X-ray images, light curves, and low-resolution spectra; two Reflection Grating Spectrometers (RGS, \citealt{XMM_rgs}), which provide high-resolution X-ray spectra of individual sources; and an Optical Monitor \citep{XMM_om} that can provide images, light curves, or low-resolution spectra in a range of optical and near-ultraviolet bands. This work focuses on the EPIC data of the sample stars.

\subsubsection{X-ray observations}

X-ray emission from stellar coronae can be studied in intricate detail with high-resolution ($R\approx 1000$) X-ray observations, for example with XMM-Newton's RGS and Chandra's \citep{Weisskopf2000} Low- and High-Energy Transmission Grating (LETG and HETG, see \citealt{Brinkman2000} and \citealt{Canizares2005}) instruments. However, such high-resolution observations are limited to the very brightest stellar X-ray sources due to the low throughput of those spectrographs. This can be an issue for future observations of exoplanet host stars, because many of those stars of interest are either too far away or too inactive to make the collection of high-resolution X-ray spectra feasible. Additionally, only stars positioned at the aimpoint of any given X-ray observation with XMM-Newton or Chandra can be used to collect high-resolution spectra with the grating spectrographs. If a star of interest has been in the field of view of an observation, but not at the aimpoint of the telescope, by default only CCD spectra of the star will be available.

I therefore perform an analysis of CCD spectra of the sample stars. CCD spectra provide lower spectral resolution (FWHM ca.\ 70~eV, i.e.\ $R\approx$ 30), but provide high S/N spectra of stars with lower X-ray flux. 
The majority of the sample stars themselves are X-ray bright enough to have decently exposed RGS or LETG/HETG spectra. Many of those high-resolution spectra have been analysed for coronal abundances in the literature, with \citet{Wood2018} presenting a recent compilation. The results derived here from CCD spectra can therefore be compared to the ones derived from high-resolution spectra; I will show in section~\ref{high_low_res_abund} that the iron abundance results agree well with each other, and X-ray CCD spectroscopy can be a useful tool to determine relative coronal abundances of some key elements.

Most of the sample stars were observed with XMM-Newton, with the exception of BF~Lyn and GJ~644. All stars observed with XMM-Newton yielded a strong detection. This is unsurprising, because the sample for the EUVE stellar atlas was based on known magnetically active stars. As such, they tend to be X-ray bright and are easily detected even in moderately short XMM-Newton exposures. 

For the two stars that were not observed with XMM-Newton, namely GJ~644 and BF~Lyn, there are archival observations with Chandra ACIS-S and ROSAT, respectively. However, it turned out that the CCD spectra extracted from those observations were not of sufficient quality to allow a detailed fit with varying coronal abundances. In the case of GJ~644, this was mainly due to the Chandra observation suffering from severe pile-up, and for BF~Lyn the ROSAT spectrum contained only a few hundred counts and the energy resolution of ROSAT is low with $\sim 0.5$\,keV (FWHM). Since the coronal abundances are a key point of this investigation, I excluded those two stars from the further analysis.

If a source was observed with several X-ray telescopes, preference was given to XMM-Newton data due to its large effective area. Many sources were observed several times. The light curves of the stars, which are automatically provided by the XMM-Newton data archive, were used for a visual check whether any large flares were occurring during a given observation. This is not unusual for stars of a high activity level. However, it is rather unlikely that a large flare occurred during the short EUVE observations as well, therefore I selected a single X-ray observation per star that did not contain any extremely large flares. If several observations without large flares were available, I chose the one with the longest non-flaring exposure time. In the rare case that only one observation was available and the observation contained a large flare, the flare was excluded from the subsequent data analysis. The selected observations are specified in Table~\ref{obs}.

\subsubsection{X-ray data analysis}

\begin{table}
\caption{Strong stellar emission lines between 200 and 504\,{\AA}.}
\label{tab:lines}
\begin{tabular}{l l l}
\hline \hline
Spectral line    & Wavelength & Peak formation\\
                 &            & temperature \\
                 & ({\AA})    & (MK) \\
\hline
Fe\sc{XIII}      & 202.04    & 1.6 \\
Fe\sc{XIII}      & 204.95    & 1.6  \\
Fe\sc{XIV}       & 211.32    & 2.0  \\
He\sc{II}        & 256.32    & 0.1 \\
Fe\sc{XIV}       & 274.20    & 2.0   \\
Fe\sc{XV}        & 284.16    & 2.0   \\
He\sc{II}        & 303.79    & 0.08  \\
Fe\sc{XVI}       & 335.41    & 2.5   \\
Fe\sc{XVI}       & 360.76    & 2.5   \\
Mg\sc{IX}        & 368.06    & 1.0   \\
Ne\sc{VII}       & 465.22    & 0.5   \\
Si\sc{XII}       & 499.41    & 2.0   \\

\hline
\end{tabular}
\end{table}

The XMM-Newton data was reduced following the standard procedures in the XMM-Newton user handbook. I used the official XMM-SAS software version 18.0.0.

To extract the spectra, I defined circular source regions, typically of 20$^{\prime\prime}$ size, centered on the target and larger background regions free of obvious sources. In cases where the target was very X-ray bright, I checked the photon event pattern distributions to test for pile-up (i.e.\ more than one photons hitting a pixel within the readout time, leading to them being registered as a single photon with higher energy), using the XMM-SAS task \texttt{epatplot}. If pile-up was present, I used an annulus source region instead of a circular one, so that the center of the point spread function (PSF) was excluded from the analysis and any pile-up related distortion in the extracted spectra was avoided. 

Source and background spectra were extracted from the PN detector if it was run in imaging mode, or from one of the lower signal-to-noise MOS detectors in case PN was run in a mode that did not yield imaging spectra. In the two cases where only Chandra or ROSAT data were available, the procedure was analogous, but using the Chandra CIAO software version 4.11 and the general-purpose Xselect software for ROSAT instead.

The spectra were fitted using Xspec version 12.10. I used a coronal plasma model, specifically several \texttt{apec} model components, to fit the stellar spectra. Absorption by the interstellar medium could be neglected because all sources are located within 30\,pc of the Sun. In order to be able to compare the plasma contributions at different temperatures across different sources, I defined a 6-temperature model grid with fixed temperatures of 1, 2, 4, 8, 16 and 32\,MK. I left the emission measure as a free parameter to fit, as well as the abundances of the most prominent elements in X-ray spectra of moderately active stars, namely oxygen, neon, and iron. The other abundances were fixed at solar values \citep{GrevesseSauval1998}.

The spectral fits directly yield a fitted flux value; I chose an energy range of 0.2-5~keV for the flux determination, as this contains the vast majority of the stellar X-ray flux even for stars with a high activity level. The flux was then converted into X-ray luminosity using Gaia DR2 distances of the stars where available, using the geometric distances of \citet{Bailer-Jones2018}. If the stars did not have parallaxes in Gaia DR2, for example because they are too bright, I used distances from Hipparcos \citep{Perryman1997}; this was the case for Altair, Procyon, $\alpha$~Cen, and GJ~644. AT~Mic is a binary star whose two components are resolved by Gaia, but not by EUVE or XMM-Newton, and I used the average distance to the two stars in the calculations.

\begin{figure*}
\includegraphics[width=0.95\columnwidth]{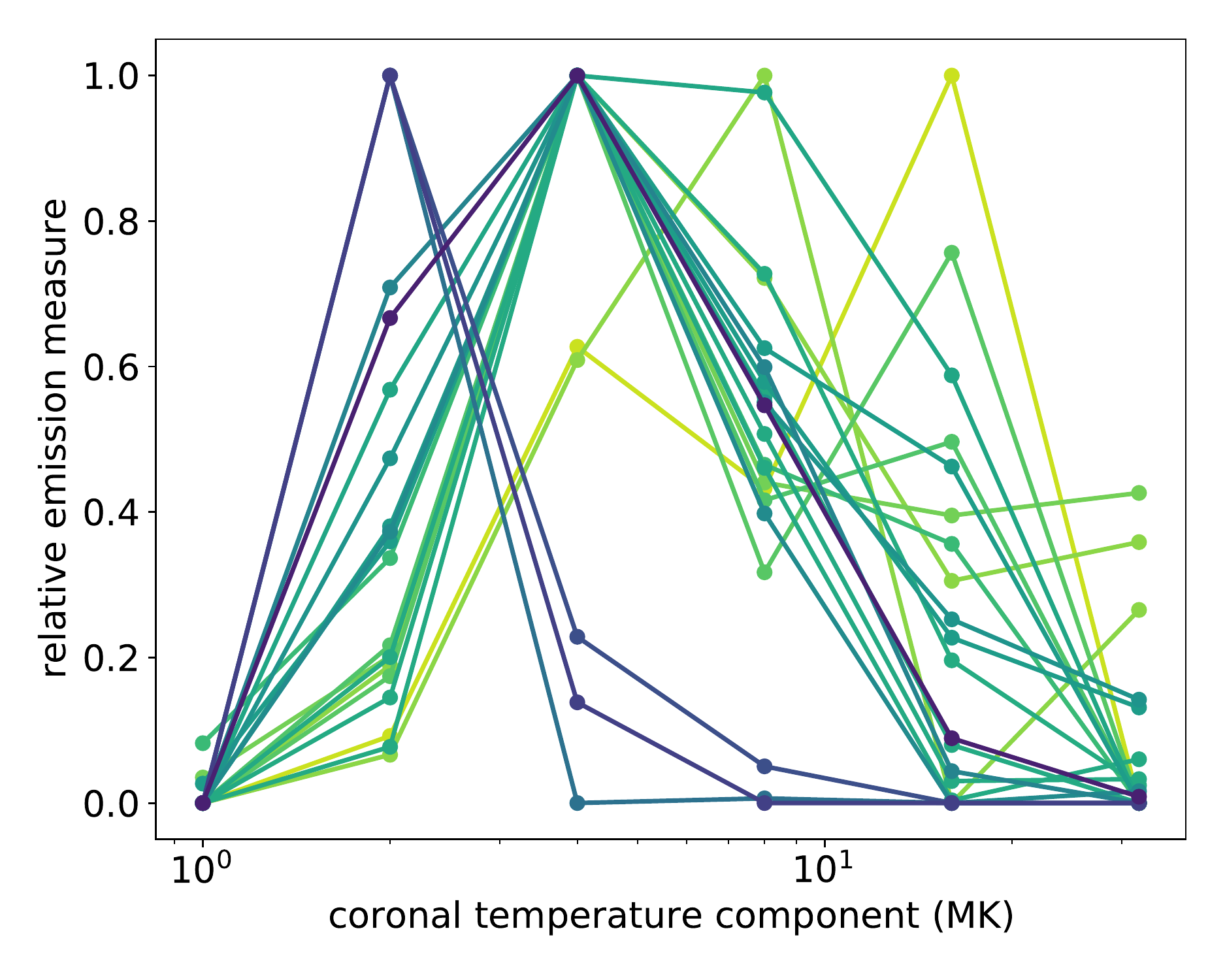}
\includegraphics[width=1.03\columnwidth]{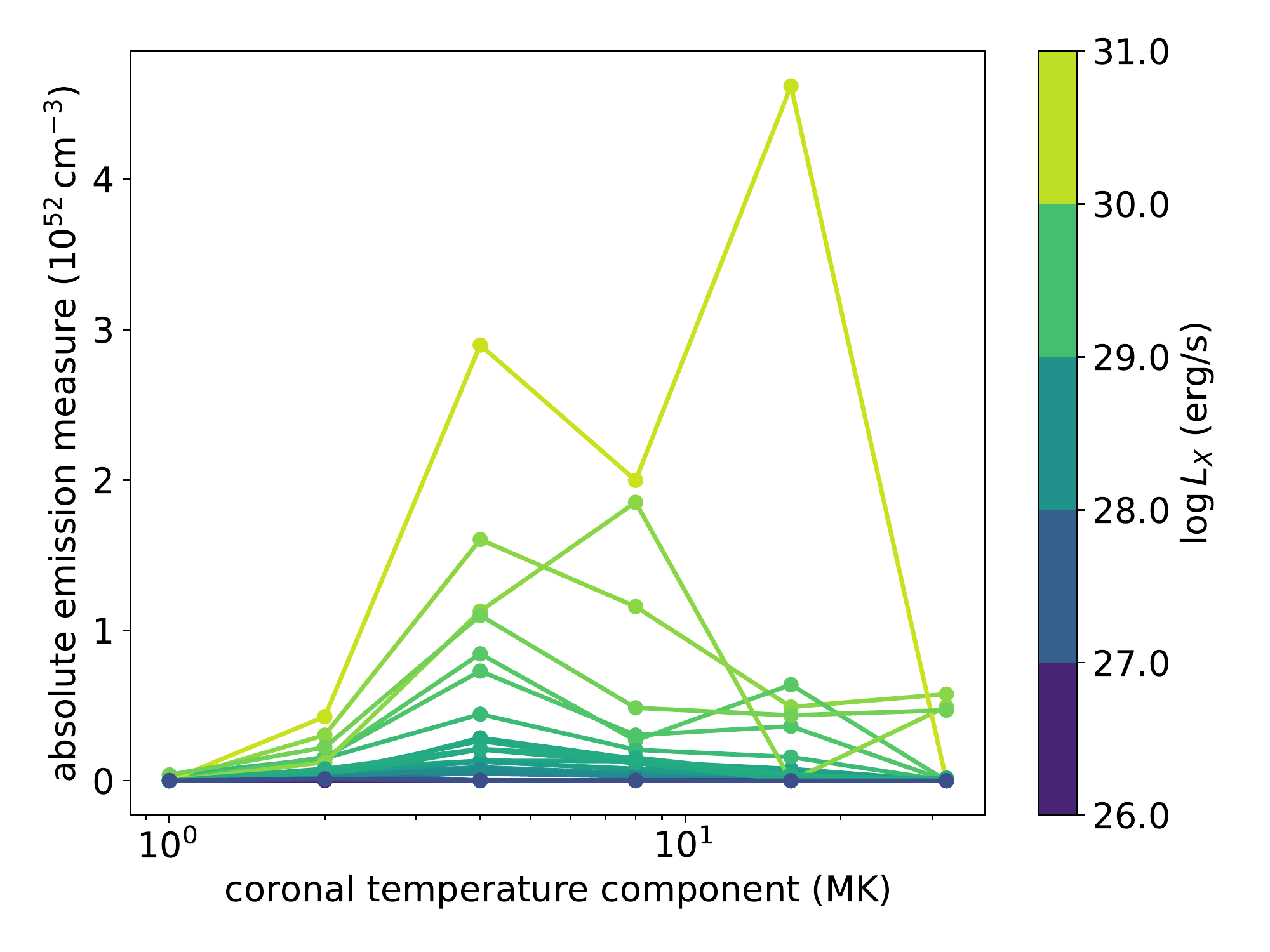}
\caption{The coronal emission measure distribution of the sample stars is shown over six temperature bins (1, 2, 4, 8, 16, and 32~MK). \textit{Left:} The relative emission measure of each star is shown (i.e.\ normalised by the largest EM component per star). The EM distributions are colour-coded by the X-ray luminosity of each star. The shift of the peak in the emission measure distribution to higher temperatures with increasing X-ray luminosity is visible. \textit{Right:} The absolute emission measure of each star is shown. The increase in total emission measure with increasing coronal temperature is visible.}
\label{fig:specfits}
\end{figure*}

\section{Results}\label{res}

\subsection{EUV iron emission lines are coronal lines}

I selected stellar spectral lines in the wavelength range of 200 to 505\,\AA\ that were detected in the highest signal-to-noise spectra of the EUVE stellar atlas \citep{Craig1997}. I then queried the ATOMDB database \citep{ATOMDB2001} for strong stellar emission lines with emissivity above $10^{-16}$ to update the wavelengths for those lines with more precise values. The resulting spectral lines, as well as their peak formation temperature from ATOMDB, are listed in Table~\ref{tab:lines}. The strongest emission line is the resonant HeI line at 303.79\,\AA, which is formed at temperatures around 80000\,K, i.e.\ temperatures typical for the stellar transition region. The other strong lines in the relevant wavelength region are formed at significantly higher temperatures; most of them have their peak formation temperature around 2~MK.

The only part of the stellar atmosphere where temperatures above a million K are found is in the stellar corona. Many of these EUV lines are therefore genuine coronal emission lines. This is relevant because these EUV lines provide a direct link to the coronal emission at X-ray wavelengths, which is easily observable with current X-ray telescopes. 

I calculated the stellar EUV fluxes by summing up the counts within 2~\AA\ of each of the lines listed in Table~\ref{tab:lines}. This way, noise from wavelength regions devoid of strong lines was  avoided in the flux determination. The resulting EUV fluxes are listed in Table~\ref{tab:sample}.

It is important to point out here that the EUV emission lines that are located close to the helium ionization threshold, i.e.\ Mg368, Ne465 and Si499, are subject to strong absorption by the interstellar medium. The observed EUVE spectra are therefore different from the spectra seen by an exoplanet orbiting the star; for the exoplanet, the irradiation in those lines will be much stronger than what we see in the absorbed spectra. I perform an estimate of the influence of those lines in section~\ref{ism_abs_quant}.

\subsection{X-ray spectral results}

The coronal iron line emission in the EUV is controlled by two factors: the available emission measure in the correct temperature range for the formation of those lines, and the coronal iron abundance. Both parameters are available from the spectral X-ray fits; the individual fitting results are listed in Table~\ref{tab:specfits}.

\subsubsection{Emission measure distributions}

The emission measure (EM) distributions fitted from the X-ray spectra of the sample stars are shown in Fig.~\ref{fig:specfits}. In the left panel, the relative EM distributions are displayed, normalised by the largest EM component of each star. As is typically seen for stellar coronae, the peak of the relative EM distribution moves to higher temperatures as the magnetic activity level and the X-ray luminosity increases (see for example \citet{Johnstone2015} and references therein).

This might give rise to the expectation that there is a "sweet spot" in coronal temperature, namely when the peak of the EM distributions is near 2~MK, where a maximum of EUV iron lines is produced. However, when looking at the absolute EMs, as shown in the right panel of Fig.~\ref{fig:specfits}, it becomes clear that the absolute emission measure at 2~MK actually rises with increasing X-ray luminosity, even if higher-temperature components become stronger in relation to it for a given star.

\subsubsection{Coronal abundances}\label{high_low_res_abund}

\begin{figure}
\includegraphics[width=\columnwidth]{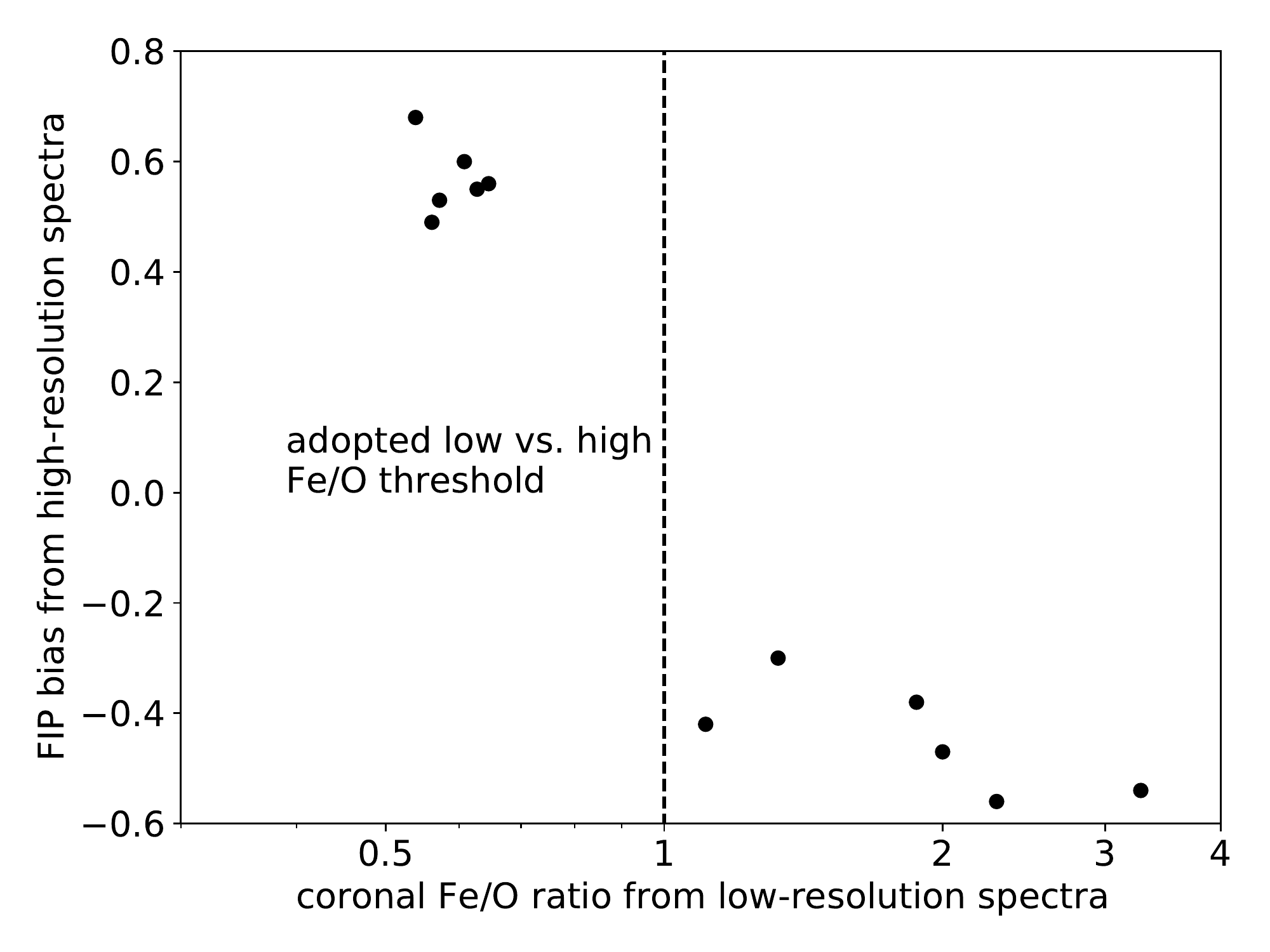}
\caption{FIP abundance bias, measured from high-resolution X-ray spectra \citep{Wood2018}, versus the iron to oxygen abundance ratio measured from low-resolution X-ray spectra in this work. A positive FIP bias is defined by a low Fe/O ratio, a negative FIP bias conversely means a high Fe/O ratio. Both measures agree in their determination of the abundance ratio, showing that low-resolution data can be used for a division of stars into low versus high relative iron abundances in stellar coronae.}
\label{fig:fip}
\end{figure}

The other quantity that influences EUV iron line emission is the coronal abundance of iron. Abundances of various elements in stellar coronae have been observed to display distinct patterns that relate to the first ionisation potential (FIP) of those elements \citep{Drake1997, Laming1999, Guedel2001}. Iron is one of the prominent elements with a low FIP. For stars with low to moderate activity, the observed coronal abundance of low-FIP elements (relative to solar photospheric abundances) is enhanced. In contrast, for stars with high magnetic activity and/or late spectral type this pattern is reversed, and an inverse FIP (iFIP) effect is observed where the elements with a high FIP such as oxygen and neon are more abundant in the corona while iron is depleted \citep{Wood2010, Liefke2008, Wood2018}. The reasons for the occurrence of this effect are thought to be rooted in the Alfv{\'e}n wave heating of the corona and the related ponderomotive force, although the detailed physical processes are not fully understood yet \citep{Laming2004, Laming2021}.

The CCD spectra analysed here have a relatively low spectral resolution. However, this is offset by the large number of photons in each spectrum which provides a high enough S/N to detect significantly non-solar ratios of iron to oxygen. To confirm that the identification of iron to oxygen ratios from CCD spectra is reliable using only high S/N CCD spectra, I compared my results to the "FIP bias" measurements by \citet{Wood2018}, who compiled coronal abundance measurements for stars from high-resolution X-ray observations. They define the FIP bias as the logarithmic ratio of several high-FIP elements (like oxygen) to several low-FIP elements (like iron). The overlap of their sample and this work consists of 12 stars, and the low-resolution O/Fe ratios derived in this work are in good agreement with the high-resolution FIP bias, as shown in Fig.~\ref{fig:fip}.

\begin{figure}
\includegraphics[width=\columnwidth]{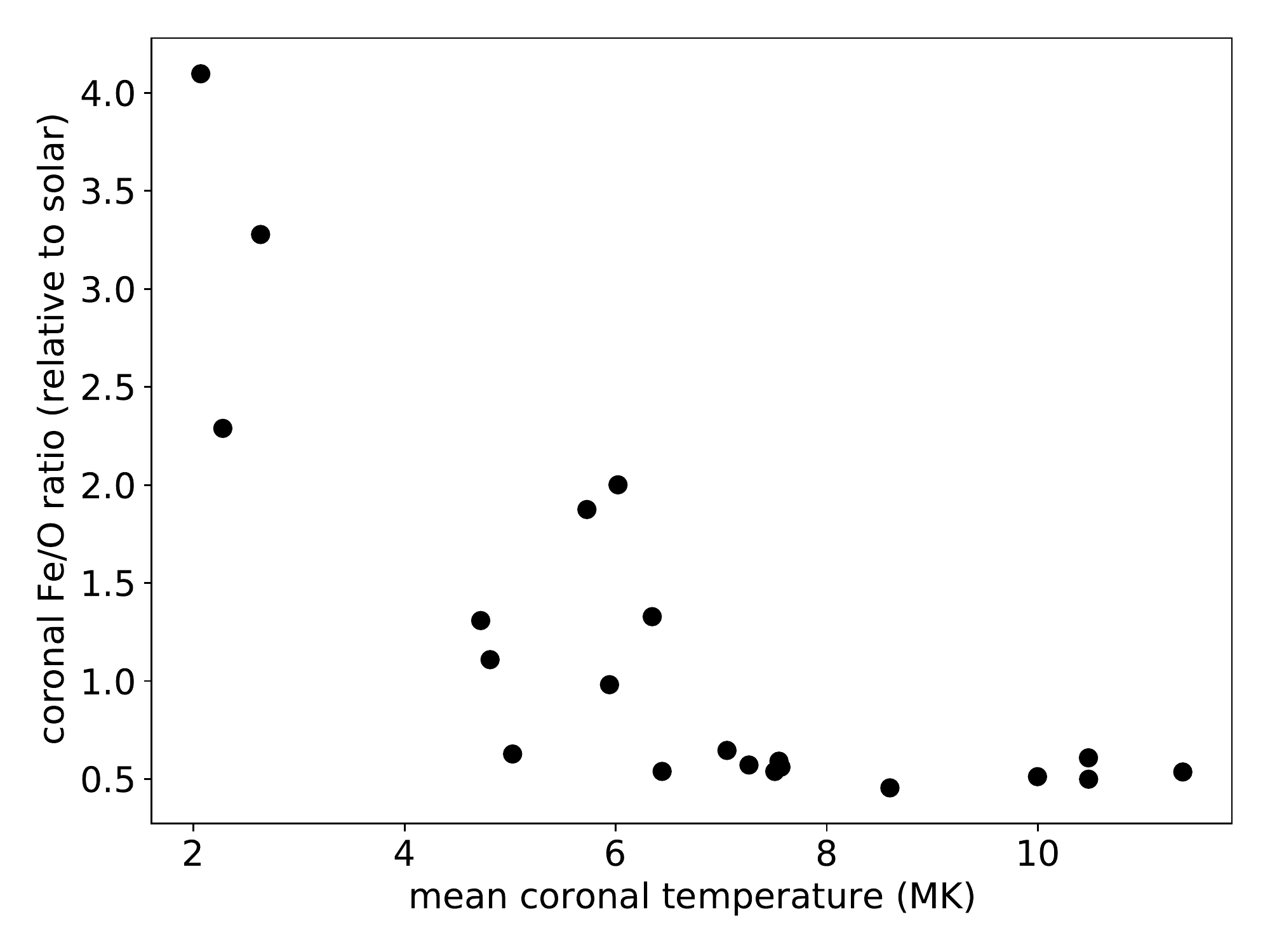}
\caption{Measured coronal iron to oxygen abundance ratios for the sample stars shown against mean coronal temperature.}
\label{fig:fip_vs_lx}
\end{figure}

The observed iron to oxygen abundance ratio in the analysed stellar sample is shown against the mean coronal temperature in Fig.~\ref{fig:fip_vs_lx}. The previously noted trend \citet{Liefke2008, Wood2018} of stars with higher activity level (and therefore higher coronal temperature) having a lower relative iron abundance is visible here is well.

\subsection{EUV spectra in comparison to coronal temperatures and abundances}\label{euv_iron}

\begin{figure}
\includegraphics[width=\columnwidth]{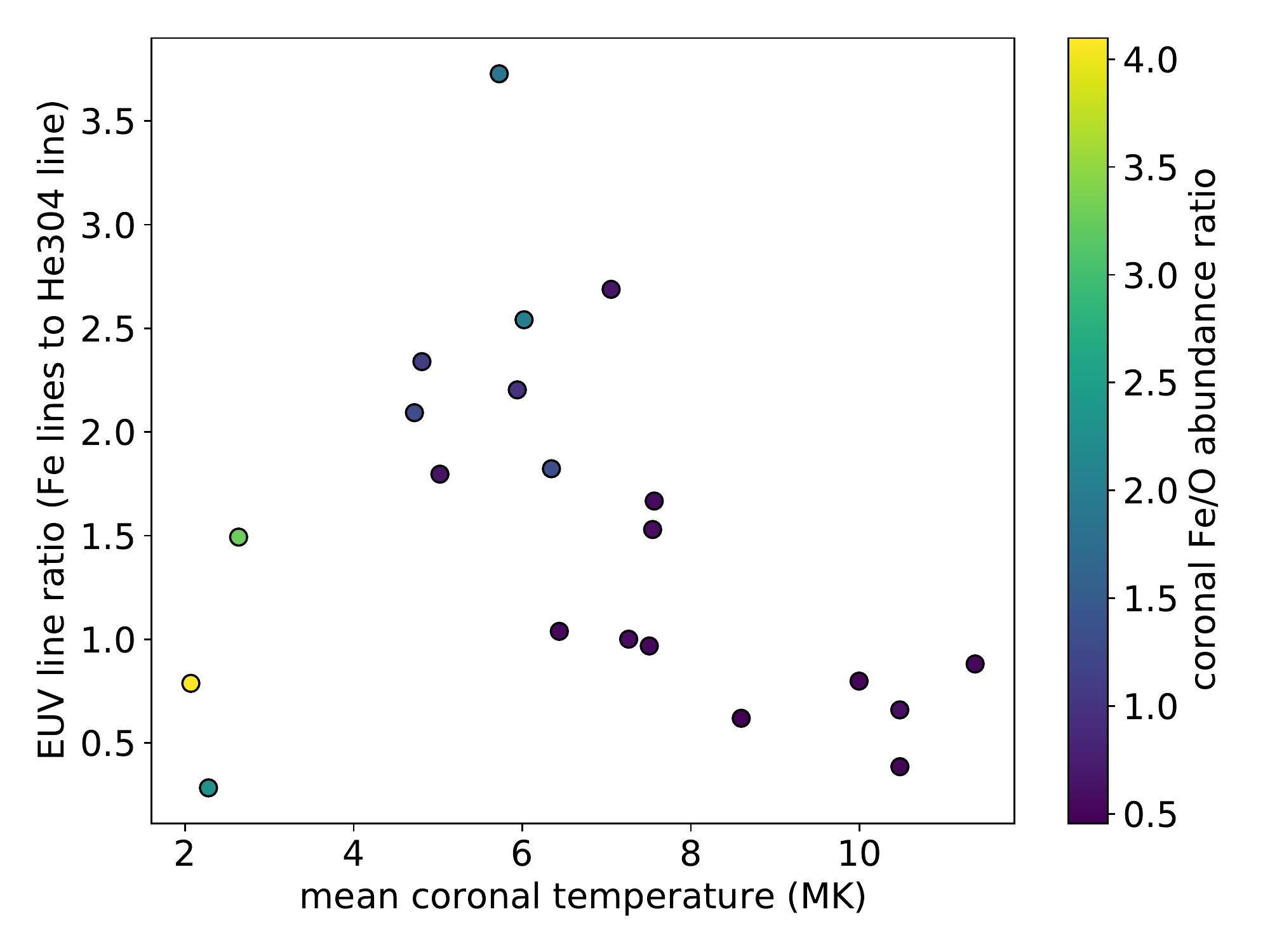}
\caption{Photon flux ratio of the coronal EUV lines versus the He\textsc{II}304 line, as a function of mean coronal temperature. The coronal iron abundance, relative to solar photospheric values, is shown by the colourbar. The decreasing iron abundance for highly active stars reduces the flux in the coronal EUV lines, many of which are iron lines.}
\label{fig:lineratio}
\end{figure}

Since the coronal emission measure around 2~MK increases with X-ray luminosity and mean coronal temperature, and the iron abundance decreases, the iron line emission in the EUV should be influenced by both effects. Indeed, the ratio of the EUV iron line from Table~\ref{tab:lines} to the He\textsc{II}304 line shows a peak over the mean coronal temperature, see Fig.~\ref{fig:lineratio}. 

This can be explained by two contributing factors: On the one hand, as the stellar activity level and the average coronal temperature increases, the filling factor of active regions on the stellar surface goes up. Therefore, even while the peak of the emission measure distribution moves to higher temperatures, the total amount of emission measure available at 2--3~MK still increases. On the other hand, at high activity levels and coronal temperatures typically seen in M dwarfs, the inverse FIP effect comes into play, where the abundance of iron and other low-FIP elements in the stellar corona decreases compared to stars of lower activity and higher mass \citep{Wood2018}. The decrease in abundance directly reduces the flux in the EUV iron lines. Note that the resulting peak in EUV iron line emission does not coincide with the peak formation temperature of those lines, which is around 2~MK, but is instead closer to 5-6~MK.

\subsection{The influence of ISM absorption on the observed spectra}\label{ism_abs_quant}

Neutral helium in the ISM is subject to the same ionisation process by the EUV lines bluewards of 504\,{\AA} as the exoplanetary atmosphere. Therefore, stellar emission lines close to the blue side of the ionization edge will undergo significant absorption by the ISM before they reach our telescopes. The solar EUV spectrum, which we can observe without strong absorption, shows some moderately strong EUV emission lines in that wavelength range, namely Ne\textsc{VII}465 and Si\textsc{XII}499 (see for example \citet{DelZanna2018, Linsky2014Chall}). In the solar spectrum the Si\textsc{XII}499 line is of roughly comparable strength to the stronger iron lines in the vicinity of He\textsc{II}304, and Ne\textsc{VII}465 line is inbetween the iron lines and the He\textsc{II}304 line in intensity \citep{DelZanna2018}.

For the observed EUVE spectra of the stars discussed here, the Ne\textsc{VII}465 and Si\textsc{XII}499 lines are practically fully absorbed by the ISM. As a rough estimate of the influence of those lines on the exoplanetary helium ionisation, I assume that the relative lines strengths are similar to the Sun. Specifically, I assume that the unabsorbed Si\textsc{XII}499 line is as strong as the Fe\textsc{xv}284 line for a given star, and the Ne\textsc{VII}465 is as strong as the He\textsc{ii}304 line. Indiviual abundances may vary, but since neon is a high-FIP element like helium and silicon is a low-FIP element like iron, the relative line strength should be more or less preserved. Since the ISM absorption is significantly weaker for the observed EUV lines between 200 and 360~{\AA}, I do not introduce any correction for those line strengths. Taking all of this into account, the estimate for the unabsorbed total EUV flux of a star between 200 and 504~{\AA} is about 1.5 times the observed absorbed flux, with a spread out to factors of 1.25 and 1.75 for individual stars in my sample.

Now, the actual ionisation in the exoplanetary atmosphere depends on the ionisation cross-section at the wavelength of a given stellar emission line, which is displayed in the top panel of Fig.~\ref{fig:spectra_cross}. One can immediately see that the ionisation cross-sections of the Si\textsc{XII}499 and Ne\textsc{VII}465 lines are about four and three times as large, respectively, as for the lines near 300~{\AA}. Therefore, they will have a relatively strong effect on the total helium ionisation in the exoplanetary atmosphere. However, since these two lines have opposite behaviours under the coronal FIP abundance effect as explained above, it turns out that the total estimated ionizing flux after including those two lines does not depend crucially on coronal abundance patterns. Rather, in comparison to only the observed (absorbed) line fluxes, the total ionizing flux is roughly a factor of three higher, but the slope of the relationship to the X-ray fluxes, which is discussed in the following section, is largely unaffected.

One caveat is that if the line fluxes of Si\textsc{XII}499 and Ne\textsc{VII}465 are very different from the solar case, then different ratios of the narrow-band EUV fluxes observed at Earth and seen by the planet may occur. In cases where a higher-precision estimate of the helium ionisation of a planet is desired, a detailed differential emission measure analysis of the star may be in order, so that the Si\textsc{XII}499 and Ne\textsc{VII}465 fluxes can be reconstructed from information gathered from other spectral lines in high-resolution X-ray observations, for example.

\subsection{The relation of X-ray and narrow-band EUV fluxes}

To estimate the production of metastable helium in exoplanet atmospheres, we are interested in the stellar narrow-band EUV flux bluewards of the helium ionisation threshold at 504~{\AA}. While the narrow-band EUV spectrum seen by the exoplanet differs slightly from what is observed by EUVE, as detailed in the previous section, I restrict myself to only the observed and therefore ISM-absorbed EUV fluxes in the following analysis, since a correction for those effects only amounts to multiplication with a constant small factor.

The relevant narrow-band EUV flux can be estimated from the observed stellar X-ray properties, either for the whole sample or for the two sub-samples split according to their coronal iron abundances. The resulting relationships can then be applied to stars where no EUV spectra are available.

\begin{figure*}
\includegraphics[height=0.35\textwidth]{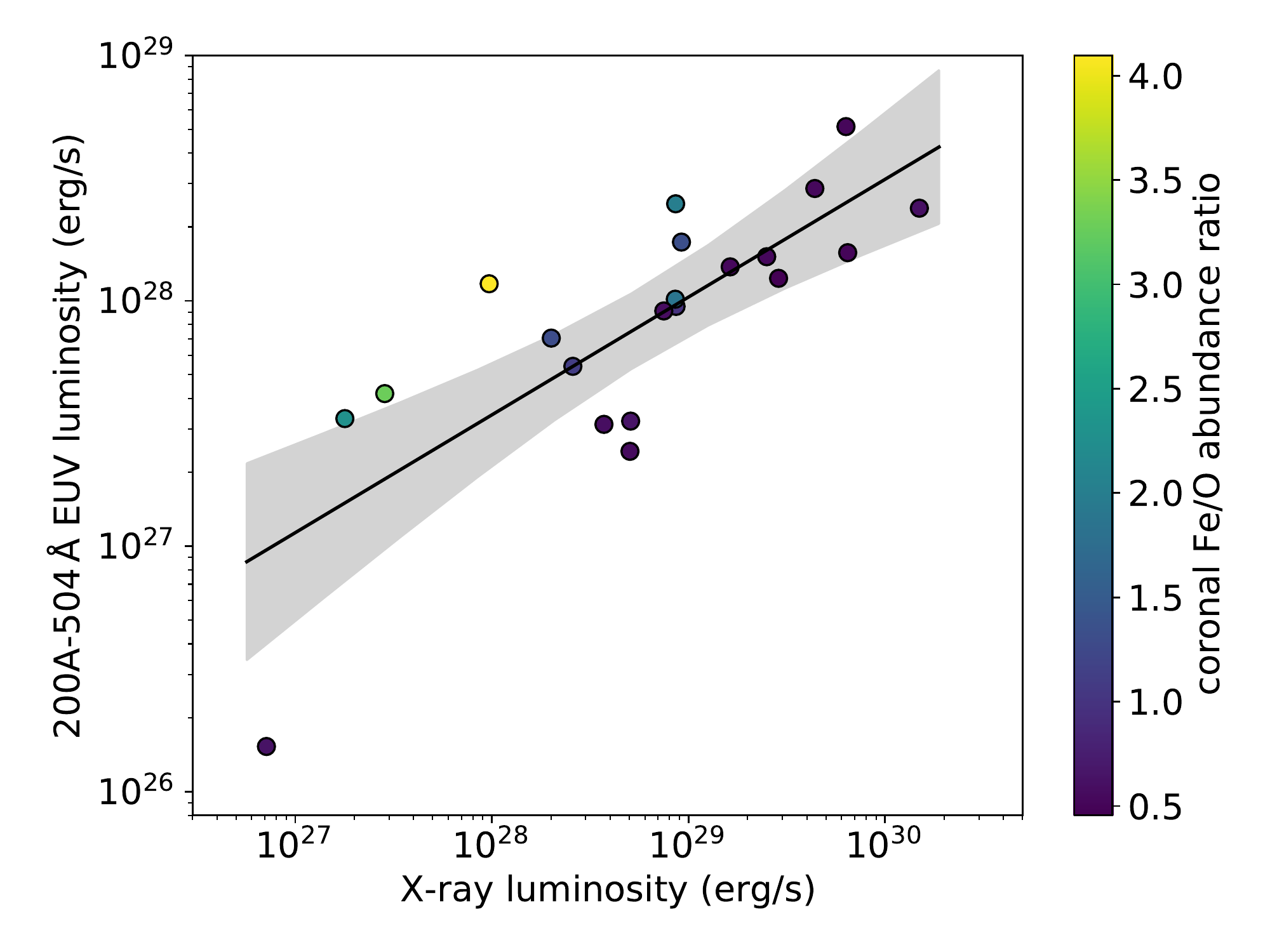}
\includegraphics[height=0.35\textwidth]{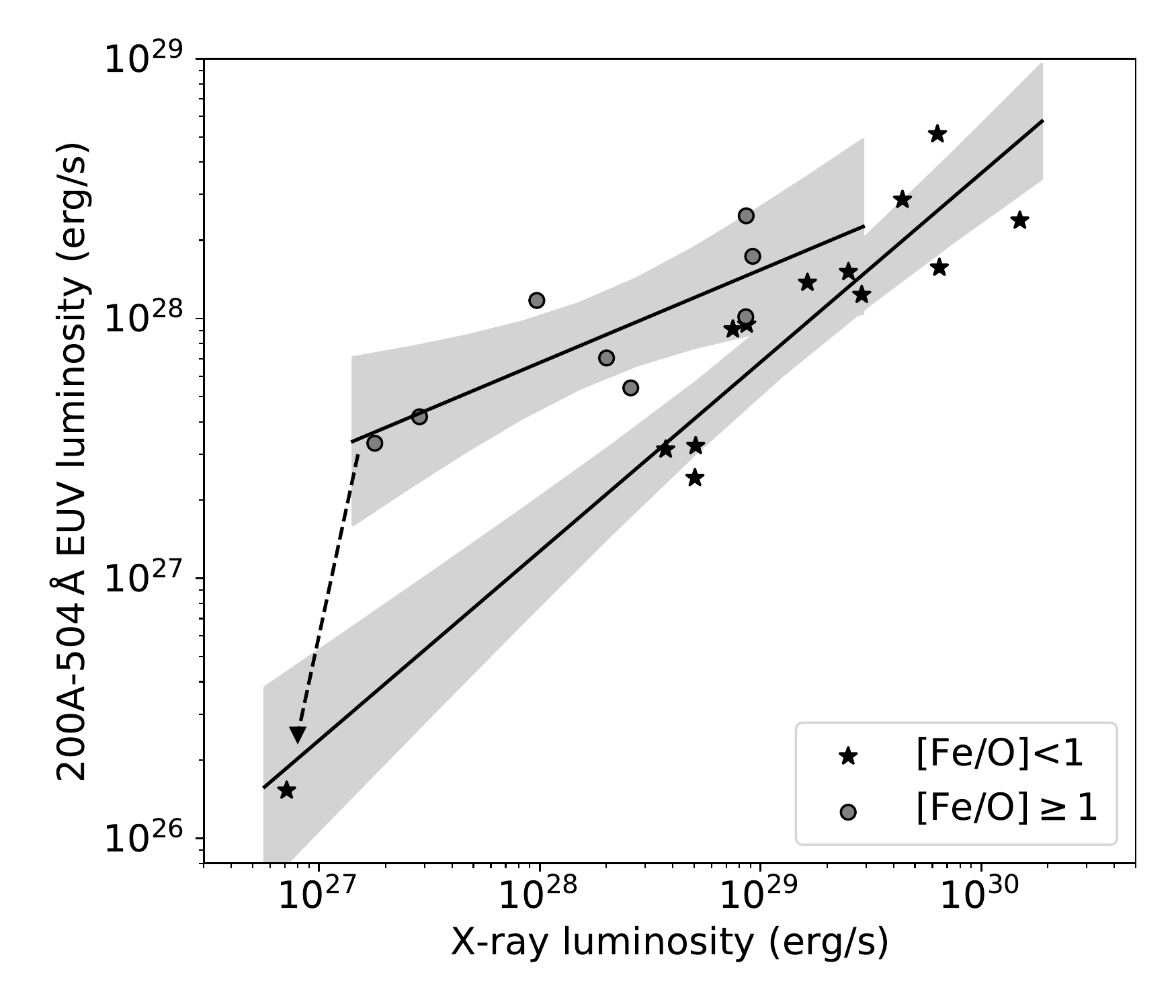}
\caption{Stellar emission in the narrow-band EUV range (200-504 \AA) versus the stellar X-ray luminosity (0.2-5 keV) for the sample stars. The full sample and the resulting fit is shown in the left panel, while the sample split by coronal iron abundance and fitting two separate relationships is shown in the right panel (see text for details).}
\label{fig:fits}
\end{figure*}

For the full sample, the relationship between the stellar narrow-band EUV luminosities in the 200--504\,\AA\ wavelength range and the stellar X-ray luminosity in the 0.2-5~keV energy range is shown in Fig.~\ref{fig:fits}, left panel. The EUV and X-ray luminosities are well correlated, as expected. I fitted the relationship between $\log L_{\mathrm{X}}$ (0.2-5~keV) and $\log L_\mathrm{200-504 \mbox{\footnotesize{\normalfont\AA}} }$ with a straight line, using the Markov Chain Monte Carlo functionality of the python package \texttt{emcee} \citep{emcee2013}.

For this whole sample, the relationship between the narrow-band EUV luminosity between 200 and 504~\AA\ and X-ray luminosity is given by

\begin{equation}
\log L_\mathrm{200-504 \mbox{\footnotesize{\normalfont\AA}} } = (0.47 \pm 0.09)\times \log \frac{L_\mathrm{X}}{10^{29}} + 28.03 \pm 0.08,
\end{equation}
which is shown together with the data in Fig.~\ref{fig:fits}.

As expected from the previous considerations in section~\ref{euv_iron}, there is some visible stratification in the data set with regard to coronal iron abundance. Specifically, stars with a high iron to oxygen abundance in their corona tend to have a higher narrow-band EUV luminosity than the stars with low coronal [Fe/O] abundance.

When splitting the sample into low-iron ([Fe/O]$<1$) and high-iron ([Fe/O]$\geq 1$) stellar coronae, two relationships with different slopes can be found:

\begin{equation}
\log L_\mathrm{low\,Fe,\, 200-504 \mbox{\footnotesize{\normalfont\AA}} } = (0.72 \pm 0.08)\times \log \frac{L_\mathrm{X}}{10^{29}} + 27.84 \pm 0.07
\end{equation}

and 

\begin{equation}
\begin{split}
\log L_\mathrm{high\,Fe,\, 200-504 \mbox{\footnotesize{\normalfont\AA}} } = (0.36 \pm 0.12)\times \log \frac{L_\mathrm{X}}{10^{29}} + 28.19 \pm 0.12\\
\mathrm{(valid~for~T_{\mathrm{cor}} \gtrsim 2~MK}).
\end{split}
\end{equation}

However, it needs to be pointed out that the relationship for the ''high Fe'' sample can only be expected to be valid for stars with (a) a high coronal iron abundance and (b) a high enough coronal temperature to form the EUV iron lines in the first place (see Fig.~\ref{fig:lineratio} and section~\ref{euv_iron}). For coronae with very low temperatures, even though the iron abundance may be high, the iron lines in the EUV will not be formed because the required ionisation state for those iron lines will not be reached in the first place. Therefore, stars with very low coronal temperature will effective behave like stars in the ''low Fe'' sample with respect to their narrowband EUV emission.

Fig.~\ref{fig:fits}, right panel, shows the relationship of the high and low iron samples. The expected drop of ''high Fe'' stars to the ''low Fe'' track is qualitatively depicted by the dashed arrow.

The two samples display different slopes and would nominally intersect around an X-ray luminosity of $\sim 10^{30}$\,erg/s. However, the sample of stars with high coronal iron abundance does not include any extremely high activity stars. This is not entirely surprising: the ''high Fe'' sample contains stars of earlier spectral type, with $\epsilon$~Eri having the latest spectral type of K2V. The ''low Fe'' sample consists mainly of M dwarfs, which remain on a high activity level much longer than more massive stars (see for example \citealt{Wright2011}). Coronal iron abundances are known to decrease with stellar mass, being typically lowest in M dwarfs \citep{Wood2018}.

\subsection{The physical meaning of non-unity slopes in the luminosity relations}

The luminosity relations derived above have slopes shallower than unity. At first glance, this is surprising: on a log-log plot, two quantities need to have different dimensionalities of growth to display a non-unity slope. If both luminosities, the 0.2-5 keV X-ray one and the EUV 200-504\,{\AA} one, both come from the stellar corona, it is not immediately obvious how one could get such different growth rates of the two luminosities. However, the key point here is the narrow temperature range over which the 200-504\,{\AA} emission is formed versus the quite broad temperature range that contributes to the X-ray emission.

Coronal plasma at a temperature of 2~MK is at the peak formation temperature of the 200-504\,{\AA} iron lines, and it also forms the very soft X-ray emission towards the low-energy end of the 0.2-5~keV X-ray emission. If we go to stars with higher X-ray luminosities, we see from Fig.~\ref{fig:specfits} (right panel) that the 2~MK emission measure increases slightly; if this were the only effect, we should see a slope of unity in the $\log L_\mathrm{200-504 \textup{\AA}}$ vs.\ $\log L_\mathrm{X}$ relation. However, we also see from Fig.~\ref{fig:specfits} that the emission measure at higher temperatures increases much more dramatically. This hotter plasma will not produce a significant increase in iron line emission in the 200-504\,{\AA} EUV range, but it will easily produce additional X-ray emission, since many lines with peak formation temperatures of 3-10~MK exist in the X-ray range, mainly bluewards of 0.5~keV. Therefore, the X-ray emission will grow much more strongly than the narrow-band EUV emission for more active stars, giving rise to a flatter-than-unity slope of the $\log L_\mathrm{200-504 \textup{\AA}}$ vs.\ $\log L_\mathrm{X}$ relation.

This also explains why the slope is flatter for the high-Fe sample than for the low-Fe one: a large fraction of the lines around 1~keV that are formed at 3-10~MK are iron lines themselves. Stars with a low coronal iron abundance will therefore display less of the aforementioned behavior, i.e.\ they will not be able to create as much excess X-ray emission relative to a given growing narrow-band EUV emission, leading to a slope closer to unity.

\section{Discussion}\label{disc}

\subsection{Helium transit depths of exoplanet atmospheres in the context of their host stars' coronae}

\begin{figure*}
\includegraphics[width=\columnwidth]{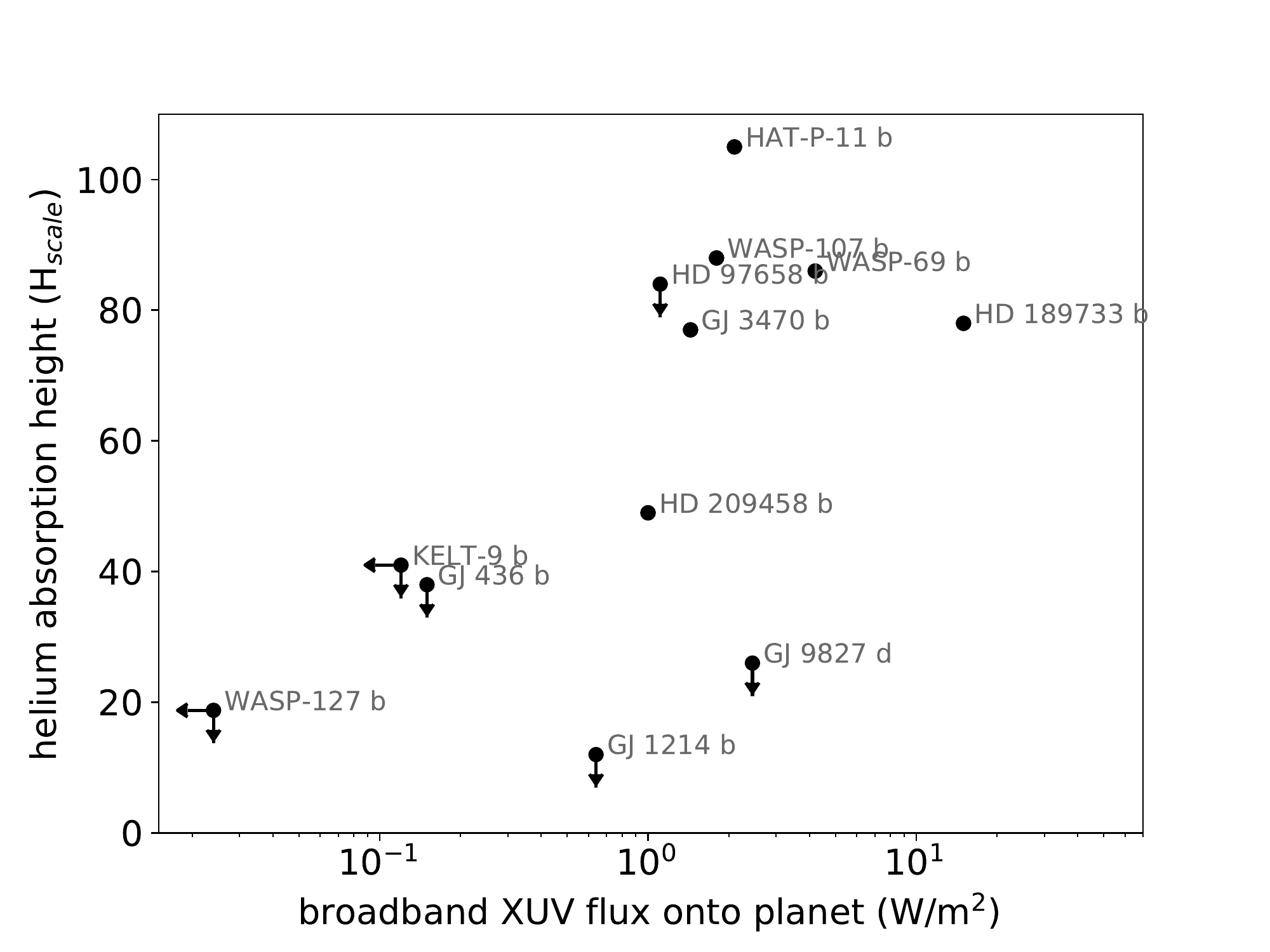}
\includegraphics[width=\columnwidth]{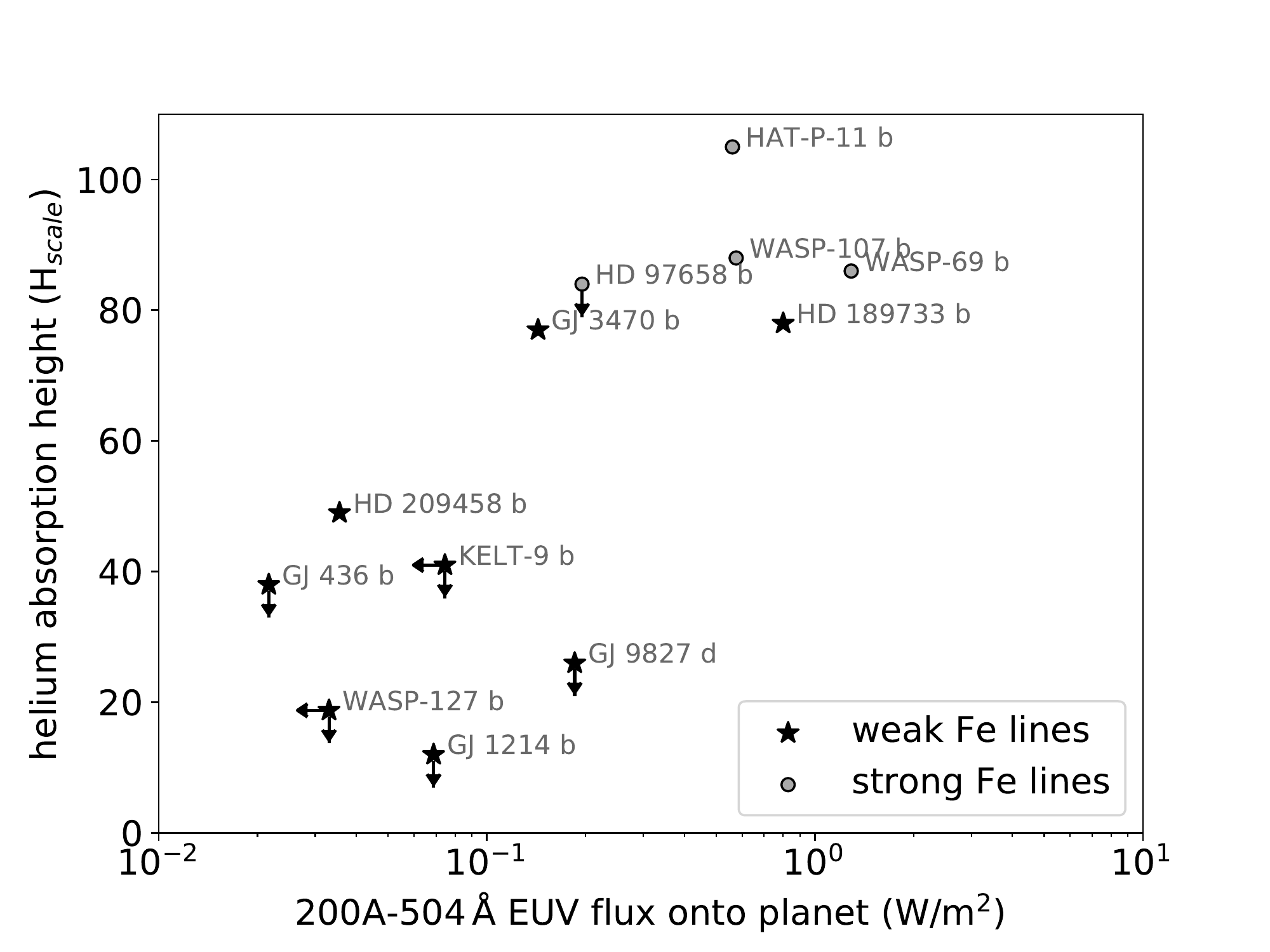}
\caption{Excess helium height of exoplanets, measured in atmospheric scale heights, as a function of exoplanet irradiation. Left panel: Fluxes on the horizontal axis are estimated broadband XUV (5-504\,\AA) fluxes from the literature. Right panel: Fluxes on the horizontal axis are estimated narrow-band EUV fluxes, assuming weak or strong iron line contributions based on stellar coronal properties.}
\label{fig:heights}
\end{figure*}

The results presented here show that the helium ionisation in exoplanetary atmospheres -- a precursor to producing atoms in the ground state of the He\textsc{I}10830 transition -- depends on the EUV irradiation of those atmospheres in a narrow wavelength region influenced by coronal iron lines. In the existing literature observed transit depths of exoplanets in the He\textsc{I}10830 line are shown in relation to their broad-band XUV irradiation. \citet{Nortmann2018} first showed that planets with strong XUV irradiation tend to display larger excess absorption in the helium triplet, which was extended by measurements performed for additional exoplanets (see for example \citealt{Alonso-Floriano2019, Kasper2020}). In those works, the XUV flux between 5 and 504 \AA\ was used, i.e.\ bluewards of the helium photoionisation edge, by applying a modified version of the scaling relationship by \cite{Sanz-Forcada2011}, or by estimating the XUV flux between 5 and 504 \AA\ from hydrogen Ly-$\alpha$ fluxes. 

I show a compilation of the published helium absorption heights in the exoplanet atmosphere versus their broadband XUV irradiation bluewards of 504~\AA\ in Fig.~\ref{fig:heights}. As can be seen, the relationship contains significant scatter. Especially the upper limits on helium absorption for exoplanets GJ~1214\,b and GJ~9827\,b fall far below the trend line established by the other exoplanets \citep{Kasper2020}.

However, it is the narrow-band EUV emission that actually drives the helium ionisation, instead of the broadband XUV emission. I estimate the narrow-band EUV irradiation of the exoplanet atmospheres from equations (2) and (3), using available information on the stellar coronae to infer whether a given stars falls into the ''high Fe'' or ''low Fe'' sample. I collected the stellar X-ray luminosities from published works, typically from \citet{Foster2021arXiv} or works on the individual host stars. In the case of GJ~9827, whose X-ray flux was not found in published works, I used the latest data from the XMM-Newton DR10 catalog \footnote{\url{http://xcatdb.unistra.fr/4xmmdr10/index.html}} with a matching radius of 10$^{\prime\prime}$ around the nominal position of the star. The XMM-Newton catalog fluxes assume an underlying power law spectrum, which is not correct for stellar coronae; I therefore use a correction factor of $F_{\mathrm{X,\,coronal}} = 0.87 F_{\mathrm{X,\,powerlaw}}$ as derived by \citet{Foster2021arXiv} to convert the DR10 catalog fluxes into coronal X-ray fluxes.

Very few of the host stars of those exoplanets have X-ray spectra of sufficient quality to determine their coronal iron abundance; only HD~189733 has been directly studied for coronal abundances so far \citep{Pillitteri2014}. I therefore rely on the results of \citet{Wood2018} to assign low coronal iron abundance to all host stars with spectral types M0 or later. Additionally, some other stars can be expected to follow the ''low Fe'' relationship as well due to very low coronal temperatures: KELT-9 is an A0V star, and if it possesses a corona at all, it can be expected to be of very low temperature around 1\,MK in analogy to the corona of $\beta$~Pictoris \citep{Guenther2012betaPic}; HD~209458 has been measured to have a low-temperature corona \citep{Czesla2017} of about 1\,MK;
and WASP-127 has been modelled to be very old with 11~Gyr \citep{Lam2017}, making it reasonable to assume its coronal temperature is of the order of 1\,MK as well. All other remaining stars were assigned to the ''high Fe'' group. The expected narrow-band EUV fluxes of the host stars were then calculated accordingly.

The resulting relationship between helium absorption heights and the narrow-band EUV irradiation is displayed in Fig.~\ref{fig:heights}, and the numerical values are listed in Table~\ref{tab:irrad}. The table gives the stellar spectral type, the planetary semi-major axis $a_\mathrm{sem}$, the published helium absorption heights in units atmospheric scale heights alongside the published broadband irradiation of the planets in the broadband XUV  ($F_\mathrm{XUV,\, pl}$); the stellar narrow-band EUV luminosity ($\log L_\mathrm{EUV}$) and the planetary irradiation in the same band ($F_\mathrm{EUV, \, pl}$) are derived from the stellar X-ray luminosity according to the two relationships given in equations 2 and 3.
The previously rather strong outlier positions of GJ~9827~d and GJ~1214~b are ameliorated. Also the position of HD~189733~b, which experiences a strong broadband XUV irradiation while having only a moderately deep helium transit, is now more in line with the other exoplanets due to the measured low coronal iron abundance of its host star.

\subsection{The nondetections of helium in sub-Neptunic exoplanet atmospheres}

A number of small exoplanets were investigated for helium absorption in their atmospheres, but even though they are subject to moderate estimated broad-band XUV irradiation from their low-mass host stars, only upper limits could be placed. This has led to questions whether Neptunic and sub-Neptunic exoplanets may have different upper atmosphere layers than Hot Jupiters, where helium absorption has been more readily observed at apparently similar XUV irradiation levels. Examples for this are the planets GJ~1214b and GJ~9827d, and a less stringent upper limit was placed on HD~97658b \citep{Kasper2020}. Also recently investigated were the planets $b$, $e$ and $f$ around Trappist-1 \citep{Krishnamurthy2021}; however, there the nondetection was less surprising since the host stars is a very old and inactive M dwarf, leading to low present-day XUV irradiation levels of the planets in the first place \citep{Wheatley2017}.

Using the EUV-X-ray relationships derived in section 3.5, it turns out that the sub-Neptunes GJ~1214b and GJ~9827d experience significantly lower levels of narrow-band EUV irradiation than the stellar $L_X$ would imply when ignoring the coronal iron lines. When one takes into account that the host stars can be expected to have only weak iron line emission (one due to its low coronal temperatue, the other due to expected low coronal iron abundance since the host star is an M dwarf), the nondetections are actually in line with the ionizing narrow-band EUV irradtion, see Fig.~\ref{fig:heights}, right panel. This indicates that those planets do not necessarily have small or helium-poor gaseous envelopes, but rather that their M dwarf host stars do not have enough coronal iron emission to excite the planetary helium into the observable metastable state.

Future observations of exoplanet host stars with decently exposed X-ray CCD spectra can be used to directly measure the relative iron abundance in the stellar coronae and constrain the expected helium absorption signal of their exoplanets further.

\begin{table*}\setlength{\tabcolsep}{4pt}
\caption{Irradiation fluxes in the XUV and the EUV band of exoplanets with measured He\textsc{I}10830 transit depths (see text for details).}
\begin{footnotesize}
\begin{tabular}{llllllllll}

\hline\hline
Name & stellar & $a_\mathrm{sem}$ & $\Delta_\mathrm{depth}$ & $F_\mathrm{XUV,\,pl}$ &  $\log L_\mathrm{X}$ & coronal iron? & $\log L_\mathrm{EUV}$ & $F_\mathrm{EUV,\,pl}$ & References $\Delta_\mathrm{depth}$\\
    & spec.\ type   & [AU] & [$H_\mathrm{scale}$] & [\flux]  & [\lumi] &  & [\lumi] & [\flux] & and $F_\mathrm{XUV,\,pl}$, $\log L_\mathrm{X}$  \\ \hline

KELT-9 b & A0V & 0.035 & <41.0 & <0.12 & <27.0 & low; cool/no corona & <26.4 & <0.07 & AF19, estimate \\
&&&&&&&&& based on spec.\ type \\
HD 209458 b & F9V & 0.047 & 49.0 & 1.0 & 26.92 & low; cool corona & 26.35 & 0.04 & AF19, Cz17 \\
WASP-127 b & G5V & 0.052 & <18.77 & <0.024 & <27.0 & low; cool corona & <26.4 & <0.03 & dS20, estimate based\\
&&&&&&&&& on old stellar age\\
HD 97658 b & K1V & 0.08 & <84.0 & 1.11 & 27.21 & high & 27.55 & 0.2 & Ka20, Ki19 \\
HD 189733 b & K2V & 0.031 & 78.0 & 15.0 & 28.3 & low; fitted abund. & 27.34 & 0.8 & AF19, Po13 \\
HAT-P-11 b & K4V & 0.053 & 105.0 & 2.1 & 27.47 & high & 27.64 & 0.56 & AF19, Fo21 \\
WASP-69 b & K5V & 0.045 & 86.0 & 4.2 & 28.11 & high & 27.87 & 1.29 & AF19, Fo21 \\
WASP-107 b & K6V & 0.055 & 88.0 & 1.8 & 27.61 & high & 27.69 & 0.58 & AF19, Fo21 \\
GJ 9827 d & K6V & 0.019 & <26.0 & 2.45 & 26.81 & low; cool corona & 26.27 & 0.19 & Ka20, this work \\
GJ 3470 b & M2V & 0.036 & 77.0 & 1.44 & 27.42 & low; M dwarf & 26.71 & 0.14 & Pa20, Fo21 \\
GJ 436 b & M3V & 0.029 & <38.0 & 0.15 & 26.04 & low; M dwarf & 25.71 & 0.02 & AF19, Fo21 \\
GJ 1214 b & M4.5V & 0.014 & <12.0 & 0.64 & 25.87 & low; M dwarf & 25.59 & 0.07 & Ka20, La14 \\

\hline
\multicolumn{10}{l}{AF19: \citet{Alonso-Floriano2019}, Cz17: \citet{Czesla2017}, dS20: \citet{dosSantos2020}, Ka20: \citet{Kasper2020},} \\
\multicolumn{10}{l}{Ki19: \citet{King2019}, Po13: \citet{Poppenhaeger2013}, Fo21: \citet{Foster2021arXiv}, Pa20: \citet{Palle2020helium}, La14: \citet{Lalitha2014}} \\

\end{tabular}
\end{footnotesize}
\label{tab:irrad}
\end{table*}

\subsection{A special role for K dwarfs as exoplanet hosts}

Four out of six K dwarf host stars in the sample of exoplanets with observed He\textsc{I}10830 transits listed in Table~\ref{tab:irrad} can be expected to have strong coronal iron emission. However, for the M dwarfs the coronal iron abundances are low even over larger samples \citep{Wood2018}. The G and F dwarfs tend to have low X-ray luminosities and low coronal temperatures due to their faster spin-down timescales and rapidly declining magnetic activity \citep{Wright2011}; therefore, unless a given host star happens to be young, their coronae are likely too cool to produce much of the relevant iron line emission. In terms of sample statistics, it is therefore more likely to find K dwarf hosts with moderately high activity as well as high coronal iron abundances than G, F or M dwarfs with strong iron emission.

K dwarfs were also identified by \citet{Oklopcic2019} as a sweet spot for helium transit observability, however their argument was based on expected helium ionisation versus depopulation of the excited level of the He\textsc{I}10830 transitions through longer-wavelength UV photons. The broad spectral shapes of G, K and M dwarf stars imply that especially M dwarfs should give rise to frequent depopulation processes. Combined with the low coronal iron emission for M dwarfs, this emphasises the important role of K dwarf hosts for future detections of He\textsc{I}10830 in exoplanet atmospheres.

\section{Conclusions}

Exoplanet transit measurements in the metastable line triplet of helium at 10830\,\AA\ have proven to be a valuable source of insight into extended exoplanet atmospheres. However, the observability of atmospheres in this line depends on the presence of exoplanetary helium in the correct state, which is typically achieved through photoionization and recombination of helium. The relevant ionization of exoplanetary helium is caused by a narrow part of the stellar emission line spectrum in the EUV that depends sensitively on the iron abundance and the temperature of the stellar corona. I have shown that the relevant exoplanet irradiation in this band can be estimated from broadband soft X-ray observations, which are readily available with current space observatories. Together with  information the coronal iron abundance a more accurate estimate of the narrow-band EUV flux can be performed, which can guide future transit observations of exoplanet atmospheres in the He\textsc{I}10839 triplet with respect to the expected absorption signal.

\section*{Acknowledgements}
I thank the reviewer, Jeff Linsky, for his illuminating comments. 

This work is based on data obtained with XMM-Newton, an ESA science mission with instruments and contributions directly funded by ESA Member States and NASA; data obtained with EUVE, a NASA mission; data obtained from the Chandra Data Archive; and data obtained from the ROSAT Data Archive of the Max-Planck-Institut f\"ur extraterrestrische Physik (MPE) at Garching, Germany. I acknowledge support from the German \emph{Leibniz-Gemeinschaft} under project number P67/2018.

\section*{Data availability}

The data in this article are available from the XMM-Newton Science Archive (\url{https://www.cosmos.esa.int/web/xmm-newton/xsa}) with the observation IDs as given in Table~\ref{tab:sample}, and as High Level Science Products from the MAST archive (\url{https://archive.stsci.edu/prepds/atlaseuve/}). The data products generated from the raw data are available upon request from the author.




\bibliographystyle{mnras}
\bibliography{katjasbib} 

\begin{thebibliography}{}
\makeatletter
\relax
\def\mn@urlcharsother{\let\do\@makeother \do\$\do\&\do\#\do\^\do\_\do\%\do\~}
\def\mn@doi{\begingroup\mn@urlcharsother \@ifnextchar [ {\mn@doi@}
  {\mn@doi@[]}}
\def\mn@doi@[#1]#2{\def\@tempa{#1}\ifx\@tempa\@empty \href
  {http://dx.doi.org/#2} {doi:#2}\else \href {http://dx.doi.org/#2} {#1}\fi
  \endgroup}
\def\mn@eprint#1#2{\mn@eprint@#1:#2::\@nil}
\def\mn@eprint@arXiv#1{\href {http://arxiv.org/abs/#1} {{\tt arXiv:#1}}}
\def\mn@eprint@dblp#1{\href {http://dblp.uni-trier.de/rec/bibtex/#1.xml}
  {dblp:#1}}
\def\mn@eprint@#1:#2:#3:#4\@nil{\def\@tempa {#1}\def\@tempb {#2}\def\@tempc
  {#3}\ifx \@tempc \@empty \let \@tempc \@tempb \let \@tempb \@tempa \fi \ifx
  \@tempb \@empty \def\@tempb {arXiv}\fi \@ifundefined
  {mn@eprint@\@tempb}{\@tempb:\@tempc}{\expandafter \expandafter \csname
  mn@eprint@\@tempb\endcsname \expandafter{\@tempc}}}

\bibitem[\protect\citeauthoryear{{Allart} et~al.,}{{Allart}
  et~al.}{2019}]{Allart2019}
{Allart} R.,  et~al., 2019, \mn@doi [\aap] {10.1051/0004-6361/201834917}, \href
  {https://ui.adsabs.harvard.edu/abs/2019A&A...623A..58A} {623, A58}

\bibitem[\protect\citeauthoryear{{Alonso-Floriano} et~al.,}{{Alonso-Floriano}
  et~al.}{2019}]{Alonso-Floriano2019}
{Alonso-Floriano} F.~J.,  et~al., 2019, \mn@doi [\aap]
  {10.1051/0004-6361/201935979}, \href
  {https://ui.adsabs.harvard.edu/abs/2019A&A...629A.110A} {629, A110}

\bibitem[\protect\citeauthoryear{{Andretta} \& {Jones}}{{Andretta} \&
  {Jones}}{1997}]{Andretta1997}
{Andretta} V.,  {Jones} H.~P.,  1997, \mn@doi [\apj] {10.1086/304760}, \href
  {https://ui.adsabs.harvard.edu/abs/1997ApJ...489..375A} {489, 375}

\bibitem[\protect\citeauthoryear{{Bailer-Jones}, {Rybizki}, {Fouesneau},
  {Mantelet}  \& {Andrae}}{{Bailer-Jones} et~al.}{2018}]{Bailer-Jones2018}
{Bailer-Jones} C.~A.~L.,  {Rybizki} J.,  {Fouesneau} M.,  {Mantelet} G.,
  {Andrae} R.,  2018, \mn@doi [\aj] {10.3847/1538-3881/aacb21}, \href
  {https://ui.adsabs.harvard.edu/abs/2018AJ....156...58B} {156, 58}

\bibitem[\protect\citeauthoryear{{Bean}, {Miller-Ricci Kempton}  \&
  {Homeier}}{{Bean} et~al.}{2010}]{Bean2010}
{Bean} J.~L.,  {Miller-Ricci Kempton} E.,   {Homeier} D.,  2010, \mn@doi [\nat]
  {10.1038/nature09596}, \href
  {https://ui.adsabs.harvard.edu/abs/2010Natur.468..669B} {468, 669}

\bibitem[\protect\citeauthoryear{{Birkby}, {de Kok}, {Brogi}, {Schwarz}  \&
  {Snellen}}{{Birkby} et~al.}{2017}]{Birkby2017}
{Birkby} J.~L.,  {de Kok} R.~J.,  {Brogi} M.,  {Schwarz} H.,   {Snellen}
  I.~A.~G.,  2017, \mn@doi [\aj] {10.3847/1538-3881/aa5c87}, \href
  {https://ui.adsabs.harvard.edu/abs/2017AJ....153..138B} {153, 138}

\bibitem[\protect\citeauthoryear{{Bourrier} et~al.,}{{Bourrier}
  et~al.}{2018}]{Bourrier2018}
{Bourrier} V.,  et~al., 2018, \mn@doi [\aap] {10.1051/0004-6361/201833675},
  \href {https://ui.adsabs.harvard.edu/abs/2018A&A...620A.147B} {620, A147}

\bibitem[\protect\citeauthoryear{{Bowyer} \& {Malina}}{{Bowyer} \&
  {Malina}}{1991}]{EUVE}
{Bowyer} S.,  {Malina} R.~F.,  1991, in {Malina} R.~F.,  {Bowyer} S.,  eds,
  Extreme Ultraviolet Astronomy. p.~397

\bibitem[\protect\citeauthoryear{{Brinkman} et~al.,}{{Brinkman}
  et~al.}{2000}]{Brinkman2000}
{Brinkman} B.~C.,  et~al., 2000, in {Truemper} J.~E.,  {Aschenbach} B.,  eds,
  Society of Photo-Optical Instrumentation Engineers (SPIE) Conference Series
  Vol. 4012, X-Ray Optics, Instruments, and Missions III. pp 81--90,
  \mn@doi{10.1117/12.391599}

\bibitem[\protect\citeauthoryear{{Canizares} et~al.,}{{Canizares}
  et~al.}{2005}]{Canizares2005}
{Canizares} C.~R.,  et~al., 2005, \mn@doi [\pasp] {10.1086/432898}, \href
  {https://ui.adsabs.harvard.edu/abs/2005PASP..117.1144C} {117, 1144}

\bibitem[\protect\citeauthoryear{{Cauley}, {Redfield}, {Jensen}, {Barman},
  {Endl}  \& {Cochran}}{{Cauley} et~al.}{2015}]{Cauley2015}
{Cauley} P.~W.,  {Redfield} S.,  {Jensen} A.~G.,  {Barman} T.,  {Endl} M.,
  {Cochran} W.~D.,  2015, \mn@doi [\apj] {10.1088/0004-637X/810/1/13}, \href
  {https://ui.adsabs.harvard.edu/abs/2015ApJ...810...13C} {810, 13}

\bibitem[\protect\citeauthoryear{{Charbonneau}, {Brown}, {Noyes}  \&
  {Gilliland}}{{Charbonneau} et~al.}{2002}]{Charbonneau2002}
{Charbonneau} D.,  {Brown} T.~M.,  {Noyes} R.~W.,   {Gilliland} R.~L.,  2002,
  \mn@doi [\apj] {10.1086/338770}, \href
  {http://adsabs.harvard.edu/abs/2002ApJ...568..377C} {568, 377}

\bibitem[\protect\citeauthoryear{{Craig} et~al.,}{{Craig}
  et~al.}{1997}]{Craig1997}
{Craig} N.,  et~al., 1997, \mn@doi [\apjs] {10.1086/313052}, \href
  {https://ui.adsabs.harvard.edu/abs/1997ApJS..113..131C} {113, 131}

\bibitem[\protect\citeauthoryear{{Czesla}, {Salz}, {Schneider}, {Mittag}  \&
  {Schmitt}}{{Czesla} et~al.}{2017}]{Czesla2017}
{Czesla} S.,  {Salz} M.,  {Schneider} P.~C.,  {Mittag} M.,   {Schmitt}
  J.~H.~M.~M.,  2017, \mn@doi [\aap] {10.1051/0004-6361/201731408}, \href
  {https://ui.adsabs.harvard.edu/abs/2017A&A...607A.101C} {607, A101}

\bibitem[\protect\citeauthoryear{{Del Zanna} \& {Mason}}{{Del Zanna} \&
  {Mason}}{2018}]{DelZanna2018}
{Del Zanna} G.,  {Mason} H.~E.,  2018, \mn@doi [Living Reviews in Solar
  Physics] {10.1007/s41116-018-0015-3}, \href
  {https://ui.adsabs.harvard.edu/abs/2018LRSP...15....5D} {15, 5}

\bibitem[\protect\citeauthoryear{{Den Herder} et~al.,}{{Den Herder}
  et~al.}{2001}]{XMM_rgs}
{Den Herder} J.~W.,  et~al., 2001, \mn@doi [\aap] {10.1051/0004-6361:20000058},
  \href {https://ui.adsabs.harvard.edu/abs/2001A&A...365L...7D} {365, L7}

\bibitem[\protect\citeauthoryear{{Dos Santos} et~al.,}{{Dos Santos}
  et~al.}{2020}]{dosSantos2020}
{Dos Santos} L.~A.,  et~al., 2020, \mn@doi [\aap]
  {10.1051/0004-6361/202038802}, \href
  {https://ui.adsabs.harvard.edu/abs/2020A&A...640A..29D} {640, A29}

\bibitem[\protect\citeauthoryear{{Drake}, {Laming}  \& {Widing}}{{Drake}
  et~al.}{1997}]{Drake1997}
{Drake} J.~J.,  {Laming} J.~M.,   {Widing} K.~G.,  1997, \mn@doi [\apj]
  {10.1086/303755}, \href
  {https://ui.adsabs.harvard.edu/abs/1997ApJ...478..403D} {478, 403}

\bibitem[\protect\citeauthoryear{{Dupree}, {Penn}  \& {Jones}}{{Dupree}
  et~al.}{1996}]{Dupree1996}
{Dupree} A.~K.,  {Penn} M.~J.,   {Jones} H.~P.,  1996, \mn@doi [\apjl]
  {10.1086/310215}, \href
  {https://ui.adsabs.harvard.edu/abs/1996ApJ...467L.121D} {467, L121}

\bibitem[\protect\citeauthoryear{{Ehrenreich} et~al.,}{{Ehrenreich}
  et~al.}{2015}]{Ehrenreich2015}
{Ehrenreich} D.,  et~al., 2015, \mn@doi [\nat] {10.1038/nature14501}, \href
  {http://adsabs.harvard.edu/abs/2015Natur.522..459E} {522, 459}

\bibitem[\protect\citeauthoryear{{Fontenla}, {Avrett}  \& {Loeser}}{{Fontenla}
  et~al.}{1990}]{Fontenla1990}
{Fontenla} J.~M.,  {Avrett} E.~H.,   {Loeser} R.,  1990, \mn@doi [\apj]
  {10.1086/168803}, \href
  {https://ui.adsabs.harvard.edu/abs/1990ApJ...355..700F} {355, 700}

\bibitem[\protect\citeauthoryear{{Fontenla}, {Avrett}  \& {Loeser}}{{Fontenla}
  et~al.}{1993}]{Fontenla1993}
{Fontenla} J.~M.,  {Avrett} E.~H.,   {Loeser} R.,  1993, \mn@doi [\apj]
  {10.1086/172443}, \href
  {https://ui.adsabs.harvard.edu/abs/1993ApJ...406..319F} {406, 319}

\bibitem[\protect\citeauthoryear{{Foreman-Mackey}, {Hogg}, {Lang}  \&
  {Goodman}}{{Foreman-Mackey} et~al.}{2013}]{emcee2013}
{Foreman-Mackey} D.,  {Hogg} D.~W.,  {Lang} D.,   {Goodman} J.,  2013, \mn@doi
  [\pasp] {10.1086/670067}, \href
  {https://ui.adsabs.harvard.edu/abs/2013PASP..125..306F} {125, 306}

\bibitem[\protect\citeauthoryear{{Foster}, {Poppenhaeger}, {Ilic}  \&
  {Schwope}}{{Foster} et~al.}{2021}]{Foster2021arXiv}
{Foster} G.,  {Poppenhaeger} K.,  {Ilic} N.,   {Schwope} A.,  2021, arXiv
  e-prints, \href {https://ui.adsabs.harvard.edu/abs/2021arXiv210614550F} {p.
  arXiv:2106.14550}

\bibitem[\protect\citeauthoryear{{Grevesse} \& {Sauval}}{{Grevesse} \&
  {Sauval}}{1998}]{GrevesseSauval1998}
{Grevesse} N.,  {Sauval} A.~J.,  1998, \mn@doi [Space Science Reviews]
  {10.1023/A:1005161325181}, \href
  {http://adsabs.harvard.edu/abs/1998SSRv...85..161G} {85, 161}

\bibitem[\protect\citeauthoryear{{G{\"u}del} et~al.,}{{G{\"u}del}
  et~al.}{2001}]{Guedel2001}
{G{\"u}del} M.,  et~al., 2001, \mn@doi [\aap] {10.1051/0004-6361:20000220},
  \href {https://ui.adsabs.harvard.edu/abs/2001A&A...365L.336G} {365, L336}

\bibitem[\protect\citeauthoryear{{Guilluy} et~al.,}{{Guilluy}
  et~al.}{2020}]{Guilluy2020}
{Guilluy} G.,  et~al., 2020, \mn@doi [\aap] {10.1051/0004-6361/202037644},
  \href {https://ui.adsabs.harvard.edu/abs/2020A&A...639A..49G} {639, A49}

\bibitem[\protect\citeauthoryear{{G{\"u}nther}, {Wolk}, {Drake}, {Lisse},
  {Robrade}  \& {Schmitt}}{{G{\"u}nther} et~al.}{2012}]{Guenther2012betaPic}
{G{\"u}nther} H.~M.,  {Wolk} S.~J.,  {Drake} J.~J.,  {Lisse} C.~M.,  {Robrade}
  J.,   {Schmitt} J.~H.~M.~M.,  2012, \mn@doi [\apj]
  {10.1088/0004-637X/750/1/78}, \href
  {http://adsabs.harvard.edu/abs/2012ApJ...750...78G} {750, 78}

\bibitem[\protect\citeauthoryear{{Jansen} et~al.,}{{Jansen}
  et~al.}{2001}]{Jansen2001}
{Jansen} F.,  et~al., 2001, \mn@doi [\aap] {10.1051/0004-6361:20000036}, \href
  {http://adsabs.harvard.edu/abs/2001A%26A...365L...1J} {365, L1}

\bibitem[\protect\citeauthoryear{{Johnstone} \& {G{\"u}del}}{{Johnstone} \&
  {G{\"u}del}}{2015}]{Johnstone2015}
{Johnstone} C.~P.,  {G{\"u}del} M.,  2015, \mn@doi [\aap]
  {10.1051/0004-6361/201425283}, \href
  {http://adsabs.harvard.edu/abs/2015A%26A...578A.129J} {578, A129}

\bibitem[\protect\citeauthoryear{{Jordan}}{{Jordan}}{1975}]{Jordan1975}
{Jordan} C.,  1975, \mn@doi [\mnras] {10.1093/mnras/170.2.429}, \href
  {https://ui.adsabs.harvard.edu/abs/1975MNRAS.170..429J} {170, 429}

\bibitem[\protect\citeauthoryear{{Kasper}, {Bean}, {Oklop{\v{c}}i{\'c}},
  {Malsky}, {Kempton}, {D{\'e}sert}, {Rogers}  \& {Mansfield}}{{Kasper}
  et~al.}{2020}]{Kasper2020}
{Kasper} D.,  {Bean} J.~L.,  {Oklop{\v{c}}i{\'c}} A.,  {Malsky} I.,  {Kempton}
  E. M.~R.,  {D{\'e}sert} J.-M.,  {Rogers} L.~A.,   {Mansfield} M.,  2020,
  \mn@doi [\aj] {10.3847/1538-3881/abbee6}, \href
  {https://ui.adsabs.harvard.edu/abs/2020AJ....160..258K} {160, 258}

\bibitem[\protect\citeauthoryear{{Keles} et~al.,}{{Keles}
  et~al.}{2020}]{Keles2020}
{Keles} E.,  et~al., 2020, \mn@doi [\mnras] {10.1093/mnras/staa2435}, \href
  {https://ui.adsabs.harvard.edu/abs/2020MNRAS.498.1023K} {498, 1023}

\bibitem[\protect\citeauthoryear{{King} \& {Wheatley}}{{King} \&
  {Wheatley}}{2021}]{King2021}
{King} G.~W.,  {Wheatley} P.~J.,  2021, \mn@doi [\mnras]
  {10.1093/mnrasl/slaa186}, \href
  {https://ui.adsabs.harvard.edu/abs/2021MNRAS.501L..28K} {501, L28}

\bibitem[\protect\citeauthoryear{{King}, {Wheatley}, {Bourrier}  \&
  {Ehrenreich}}{{King} et~al.}{2019}]{King2019}
{King} G.~W.,  {Wheatley} P.~J.,  {Bourrier} V.,   {Ehrenreich} D.,  2019,
  \mn@doi [\mnras] {10.1093/mnrasl/slz003}, \href
  {https://ui.adsabs.harvard.edu/abs/2019MNRAS.484L..49K} {484, L49}

\bibitem[\protect\citeauthoryear{{Kreidberg} et~al.,}{{Kreidberg}
  et~al.}{2014}]{Kreidberg2014}
{Kreidberg} L.,  et~al., 2014, \mn@doi [\nat] {10.1038/nature12888}, \href
  {https://ui.adsabs.harvard.edu/abs/2014Natur.505...69K} {505, 69}

\bibitem[\protect\citeauthoryear{{Krishnamurthy} et~al.,}{{Krishnamurthy}
  et~al.}{2021}]{Krishnamurthy2021}
{Krishnamurthy} V.,  et~al., 2021, \mn@doi [\aj] {10.3847/1538-3881/ac0d57},
  \href {https://ui.adsabs.harvard.edu/abs/2021AJ....162...82K} {162, 82}

\bibitem[\protect\citeauthoryear{{Kulow}, {France}, {Linsky}  \&
  {Loyd}}{{Kulow} et~al.}{2014}]{Kulow2014}
{Kulow} J.~R.,  {France} K.,  {Linsky} J.,   {Loyd} R.~O.~P.,  2014, \mn@doi
  [\apj] {10.1088/0004-637X/786/2/132}, \href
  {http://adsabs.harvard.edu/abs/2014ApJ...786..132K} {786, 132}

\bibitem[\protect\citeauthoryear{{Lalitha}, {Poppenhaeger}, {Singh}, {Czesla}
  \& {Schmitt}}{{Lalitha} et~al.}{2014}]{Lalitha2014}
{Lalitha} S.,  {Poppenhaeger} K.,  {Singh} K.~P.,  {Czesla} S.,   {Schmitt}
  J.~H.~M.~M.,  2014, \mn@doi [\apjl] {10.1088/2041-8205/790/1/L11}, \href
  {http://adsabs.harvard.edu/abs/2014ApJ...790L..11L} {790, L11}

\bibitem[\protect\citeauthoryear{{Lam} et~al.,}{{Lam} et~al.}{2017}]{Lam2017}
{Lam} K.~W.~F.,  et~al., 2017, \mn@doi [\aap] {10.1051/0004-6361/201629403},
  \href {https://ui.adsabs.harvard.edu/abs/2017A&A...599A...3L} {599, A3}

\bibitem[\protect\citeauthoryear{{Laming}}{{Laming}}{2004}]{Laming2004}
{Laming} J.~M.,  2004, \mn@doi [\apj] {10.1086/423780}, \href
  {https://ui.adsabs.harvard.edu/abs/2004ApJ...614.1063L} {614, 1063}

\bibitem[\protect\citeauthoryear{{Laming}}{{Laming}}{2021}]{Laming2021}
{Laming} J.~M.,  2021, \mn@doi [\apj] {10.3847/1538-4357/abd9c3}, \href
  {https://ui.adsabs.harvard.edu/abs/2021ApJ...909...17L} {909, 17}

\bibitem[\protect\citeauthoryear{{Laming} \& {Drake}}{{Laming} \&
  {Drake}}{1999}]{Laming1999}
{Laming} J.~M.,  {Drake} J.~J.,  1999, \mn@doi [\apj] {10.1086/307112}, \href
  {https://ui.adsabs.harvard.edu/abs/1999ApJ...516..324L} {516, 324}

\bibitem[\protect\citeauthoryear{{Lecavelier Des Etangs} et~al.,}{{Lecavelier
  Des Etangs} et~al.}{2010}]{Lecavelier2010}
{Lecavelier Des Etangs} A.,  et~al., 2010, \mn@doi [\aap]
  {10.1051/0004-6361/200913347}, \href
  {http://adsabs.harvard.edu/abs/2010A%26A...514A..72L} {514, A72}

\bibitem[\protect\citeauthoryear{{Libbrecht}, {de la Cruz Rodr{\'\i}guez},
  {Danilovic}, {Leenaarts}  \& {Pazira}}{{Libbrecht}
  et~al.}{2019}]{Libbrecht2019}
{Libbrecht} T.,  {de la Cruz Rodr{\'\i}guez} J.,  {Danilovic} S.,  {Leenaarts}
  J.,   {Pazira} H.,  2019, \mn@doi [\aap] {10.1051/0004-6361/201833610}, \href
  {https://ui.adsabs.harvard.edu/abs/2019A&A...621A..35L} {621, A35}

\bibitem[\protect\citeauthoryear{{Liefke}, {Ness}, {Schmitt}  \&
  {Maggio}}{{Liefke} et~al.}{2008}]{Liefke2008}
{Liefke} C.,  {Ness} J.~U.,  {Schmitt} J.~H.~M.~M.,   {Maggio} A.,  2008,
  \mn@doi [\aap] {10.1051/0004-6361:200810054}, \href
  {https://ui.adsabs.harvard.edu/abs/2008A&A...491..859L} {491, 859}

\bibitem[\protect\citeauthoryear{{Linsky}}{{Linsky}}{2014}]{Linsky2014Chall}
{Linsky} J.,  2014, \mn@doi [Challenges] {10.3390/challe5020351}, \href
  {https://ui.adsabs.harvard.edu/abs/2014Chall...5..351L} {5, 351}

\bibitem[\protect\citeauthoryear{{Linsky}}{{Linsky}}{2017}]{Linsky2017}
{Linsky} J.~L.,  2017, \mn@doi [\araa] {10.1146/annurev-astro-091916-055327},
  \href {https://ui.adsabs.harvard.edu/abs/2017ARA&A..55..159L} {55, 159}

\bibitem[\protect\citeauthoryear{{Llama}, {Wood}, {Jardine}, {Vidotto},
  {Helling}, {Fossati}  \& {Haswell}}{{Llama} et~al.}{2011}]{Llama2011}
{Llama} J.,  {Wood} K.,  {Jardine} M.,  {Vidotto} A.~A.,  {Helling} C.,
  {Fossati} L.,   {Haswell} C.~A.,  2011, \mn@doi [\mnras]
  {10.1111/j.1745-3933.2011.01093.x}, \href
  {https://ui.adsabs.harvard.edu/abs/2011MNRAS.416L..41L} {416, L41}

\bibitem[\protect\citeauthoryear{{Mansfield} et~al.,}{{Mansfield}
  et~al.}{2018}]{Mansfield2018}
{Mansfield} M.,  et~al., 2018, \mn@doi [\apjl] {10.3847/2041-8213/aaf166},
  \href {https://ui.adsabs.harvard.edu/abs/2018ApJ...868L..34M} {868, L34}

\bibitem[\protect\citeauthoryear{{Mason} et~al.,}{{Mason}
  et~al.}{2001}]{XMM_om}
{Mason} K.~O.,  et~al., 2001, \mn@doi [\aap] {10.1051/0004-6361:20000044},
  \href {https://ui.adsabs.harvard.edu/abs/2001A&A...365L..36M} {365, L36}

\bibitem[\protect\citeauthoryear{{Nortmann} et~al.,}{{Nortmann}
  et~al.}{2018}]{Nortmann2018}
{Nortmann} L.,  et~al., 2018, \mn@doi [Science] {10.1126/science.aat5348},
  \href {https://ui.adsabs.harvard.edu/abs/2018Sci...362.1388N} {362, 1388}

\bibitem[\protect\citeauthoryear{{Oklop{\v{c}}i{\'c}}}{{Oklop{\v{c}}i{\'c}}}{2019}]{Oklopcic2019}
{Oklop{\v{c}}i{\'c}} A.,  2019, \mn@doi [\apj] {10.3847/1538-4357/ab2f7f},
  \href {https://ui.adsabs.harvard.edu/abs/2019ApJ...881..133O} {881, 133}

\bibitem[\protect\citeauthoryear{{Oklop{\v{c}}i{\'c}} \&
  {Hirata}}{{Oklop{\v{c}}i{\'c}} \& {Hirata}}{2018}]{Oklopcic2018}
{Oklop{\v{c}}i{\'c}} A.,  {Hirata} C.~M.,  2018, \mn@doi [\apjl]
  {10.3847/2041-8213/aaada9}, \href
  {https://ui.adsabs.harvard.edu/abs/2018ApJ...855L..11O} {855, L11}

\bibitem[\protect\citeauthoryear{{Palle} et~al.,}{{Palle}
  et~al.}{2020}]{Palle2020helium}
{Palle} E.,  et~al., 2020, \mn@doi [\aap] {10.1051/0004-6361/202037719}, \href
  {https://ui.adsabs.harvard.edu/abs/2020A&A...638A..61P} {638, A61}

\bibitem[\protect\citeauthoryear{{Paragas} et~al.,}{{Paragas}
  et~al.}{2021}]{Paragas2021}
{Paragas} K.,  et~al., 2021, \mn@doi [\apjl] {10.3847/2041-8213/abe706}, \href
  {https://ui.adsabs.harvard.edu/abs/2021ApJ...909L..10P} {909, L10}

\bibitem[\protect\citeauthoryear{{Perryman} et~al.,}{{Perryman}
  et~al.}{1997}]{Perryman1997}
{Perryman} M.~A.~C.,  et~al., 1997, \aap, \href
  {https://ui.adsabs.harvard.edu/abs/1997A&A...323L..49P} {500, 501}

\bibitem[\protect\citeauthoryear{{Pillitteri}, {Wolk}, {Lopez-Santiago},
  {Guenther}, {Sciortino}, {Cohen}, {Kashyap}  \& {Drake}}{{Pillitteri}
  et~al.}{2014}]{Pillitteri2014}
{Pillitteri} I.,  {Wolk} S.~J.,  {Lopez-Santiago} J.,  {Guenther} H.~M.,
  {Sciortino} S.,  {Cohen} O.,  {Kashyap} V.,   {Drake} J.~J.,  2014, \apj,
  785, 145+

\bibitem[\protect\citeauthoryear{{Poppenhaeger}, {Schmitt}  \&
  {Wolk}}{{Poppenhaeger} et~al.}{2013}]{Poppenhaeger2013}
{Poppenhaeger} K.,  {Schmitt} J.~H.~M.~M.,   {Wolk} S.~J.,  2013, \mn@doi
  [\apj] {10.1088/0004-637X/773/1/62}, \href
  {http://adsabs.harvard.edu/abs/2013ApJ...773...62P} {773, 62}

\bibitem[\protect\citeauthoryear{{Sanz-Forcada} \& {Dupree}}{{Sanz-Forcada} \&
  {Dupree}}{2008}]{Sanz-Forcada2008}
{Sanz-Forcada} J.,  {Dupree} A.~K.,  2008, \mn@doi [\aap]
  {10.1051/0004-6361:20078501}, \href
  {https://ui.adsabs.harvard.edu/abs/2008A&A...488..715S} {488, 715}

\bibitem[\protect\citeauthoryear{{Sanz-Forcada}, {Micela}, {Ribas}, {Pollock},
  {Eiroa}, {Velasco}, {Solano}  \& {Garc{\'{\i}}a-{\'A}lvarez}}{{Sanz-Forcada}
  et~al.}{2011}]{Sanz-Forcada2011}
{Sanz-Forcada} J.,  {Micela} G.,  {Ribas} I.,  {Pollock} A.~M.~T.,  {Eiroa} C.,
   {Velasco} A.,  {Solano} E.,   {Garc{\'{\i}}a-{\'A}lvarez} D.,  2011, \mn@doi
  [\aap] {10.1051/0004-6361/201116594}, \href
  {http://cdsads.u-strasbg.fr/abs/2011A%26A...532A...6S} {532, A6+}

\bibitem[\protect\citeauthoryear{{Seager} \& {Sasselov}}{{Seager} \&
  {Sasselov}}{2000}]{Seager2000}
{Seager} S.,  {Sasselov} D.~D.,  2000, \mn@doi [\apj] {10.1086/309088}, \href
  {https://ui.adsabs.harvard.edu/abs/2000ApJ...537..916S} {537, 916}

\bibitem[\protect\citeauthoryear{{Smith}, {Brickhouse}, {Liedahl}  \&
  {Raymond}}{{Smith} et~al.}{2001}]{ATOMDB2001}
{Smith} R.~K.,  {Brickhouse} N.~S.,  {Liedahl} D.~A.,   {Raymond} J.~C.,  2001,
  \mn@doi [\apjl] {10.1086/322992}, \href
  {https://ui.adsabs.harvard.edu/abs/2001ApJ...556L..91S} {556, L91}

\bibitem[\protect\citeauthoryear{{Snellen}, {Brandl}, {de Kok}, {Brogi},
  {Birkby}  \& {Schwarz}}{{Snellen} et~al.}{2014}]{Snellen2014}
{Snellen} I. A.~G.,  {Brandl} B.~R.,  {de Kok} R.~J.,  {Brogi} M.,  {Birkby}
  J.,   {Schwarz} H.,  2014, \mn@doi [\nat] {10.1038/nature13253}, \href
  {https://ui.adsabs.harvard.edu/abs/2014Natur.509...63S} {509, 63}

\bibitem[\protect\citeauthoryear{{Str{\"u}der} et~al.,}{{Str{\"u}der}
  et~al.}{2001}]{EPIC_pn}
{Str{\"u}der} L.,  et~al., 2001, \mn@doi [\aap] {10.1051/0004-6361:20000066},
  \href {https://ui.adsabs.harvard.edu/abs/2001A&A...365L..18S} {365, L18}

\bibitem[\protect\citeauthoryear{{Tinetti} et~al.,}{{Tinetti}
  et~al.}{2007}]{Tinetti2007}
{Tinetti} G.,  et~al., 2007, \mn@doi [\nat] {10.1038/nature06002}, \href
  {http://adsabs.harvard.edu/abs/2007Natur.448..169T} {448, 169}

\bibitem[\protect\citeauthoryear{{Turner} et~al.,}{{Turner}
  et~al.}{2001}]{EPIC_mos}
{Turner} M.~J.~L.,  et~al., 2001, \mn@doi [\aap] {10.1051/0004-6361:20000087},
  \href {https://ui.adsabs.harvard.edu/abs/2001A&A...365L..27T} {365, L27}

\bibitem[\protect\citeauthoryear{{Vernazza}, {Avrett}  \& {Loeser}}{{Vernazza}
  et~al.}{1981}]{Vernazza1981}
{Vernazza} J.~E.,  {Avrett} E.~H.,   {Loeser} R.,  1981, \mn@doi [\apjs]
  {10.1086/190731}, \href
  {https://ui.adsabs.harvard.edu/abs/1981ApJS...45..635V} {45, 635}

\bibitem[\protect\citeauthoryear{{Vidal-Madjar}, {Lecavelier des Etangs},
  {D{\'e}sert}, {Ballester}, {Ferlet}, {H{\'e}brard}  \&
  {Mayor}}{{Vidal-Madjar} et~al.}{2003}]{Vidal-Madjar2003}
{Vidal-Madjar} A.,  {Lecavelier des Etangs} A.,  {D{\'e}sert} J.,  {Ballester}
  G.~E.,  {Ferlet} R.,  {H{\'e}brard} G.,   {Mayor} M.,  2003, \nat, \href
  {http://adsabs.harvard.edu/abs/2003Natur.422..143V} {422, 143}

\bibitem[\protect\citeauthoryear{{Weisskopf}, {Tananbaum}, {Van Speybroeck}  \&
  {O'Dell}}{{Weisskopf} et~al.}{2000}]{Weisskopf2000}
{Weisskopf} M.~C.,  {Tananbaum} H.~D.,  {Van Speybroeck} L.~P.,   {O'Dell}
  S.~L.,  2000, in {J.~E.~Tr{\"u}mper \& B.~Aschenbach} ed.,  Society of
  Photo-Optical Instrumentation Engineers (SPIE) Conference Series Vol. 4012,
  Society of Photo-Optical Instrumentation Engineers (SPIE) Conference Series.
  pp 2--16 (\mn@eprint {} {arXiv:astro-ph/0004127})

\bibitem[\protect\citeauthoryear{{Wheatley}, {Louden}, {Bourrier}, {Ehrenreich}
   \& {Gillon}}{{Wheatley} et~al.}{2017}]{Wheatley2017}
{Wheatley} P.~J.,  {Louden} T.,  {Bourrier} V.,  {Ehrenreich} D.,   {Gillon}
  M.,  2017, \mn@doi [\mnras] {10.1093/mnrasl/slw192}, \href
  {https://ui.adsabs.harvard.edu/abs/2017MNRAS.465L..74W} {465, L74}

\bibitem[\protect\citeauthoryear{{Wood} \& {Linsky}}{{Wood} \&
  {Linsky}}{2010}]{Wood2010}
{Wood} B.~E.,  {Linsky} J.~L.,  2010, \mn@doi [\apj]
  {10.1088/0004-637X/717/2/1279}, \href
  {https://ui.adsabs.harvard.edu/abs/2010ApJ...717.1279W} {717, 1279}

\bibitem[\protect\citeauthoryear{{Wood}, {Laming}, {Warren}  \&
  {Poppenhaeger}}{{Wood} et~al.}{2018}]{Wood2018}
{Wood} B.~E.,  {Laming} J.~M.,  {Warren} H.~P.,   {Poppenhaeger} K.,  2018,
  \mn@doi [\apj] {10.3847/1538-4357/aaccf6}, \href
  {https://ui.adsabs.harvard.edu/abs/2018ApJ...862...66W} {862, 66}

\bibitem[\protect\citeauthoryear{{Wright}, {Drake}, {Mamajek}  \&
  {Henry}}{{Wright} et~al.}{2011}]{Wright2011}
{Wright} N.~J.,  {Drake} J.~J.,  {Mamajek} E.~E.,   {Henry} G.~W.,  2011,
  \mn@doi [\apj] {10.1088/0004-637X/743/1/48}, \href
  {http://adsabs.harvard.edu/abs/2011ApJ...743...48W} {743, 48}

\bibitem[\protect\citeauthoryear{{Yan}, {Sadeghpour}  \& {Dalgarno}}{{Yan}
  et~al.}{1998}]{Yan1998}
{Yan} M.,  {Sadeghpour} H.~R.,   {Dalgarno} A.,  1998, \mn@doi [\apj]
  {10.1086/305420}, \href
  {https://ui.adsabs.harvard.edu/abs/1998ApJ...496.1044Y} {496, 1044}

\bibitem[\protect\citeauthoryear{{Zarro} \& {Zirin}}{{Zarro} \&
  {Zirin}}{1986}]{Zarro1986}
{Zarro} D.~M.,  {Zirin} H.,  1986, \mn@doi [\apj] {10.1086/164170}, \href
  {https://ui.adsabs.harvard.edu/abs/1986ApJ...304..365Z} {304, 365}

\makeatother
\end{thebibliography}



\section{Additional figures}

\begin{figure*}
\includegraphics[height=0.95\textheight]{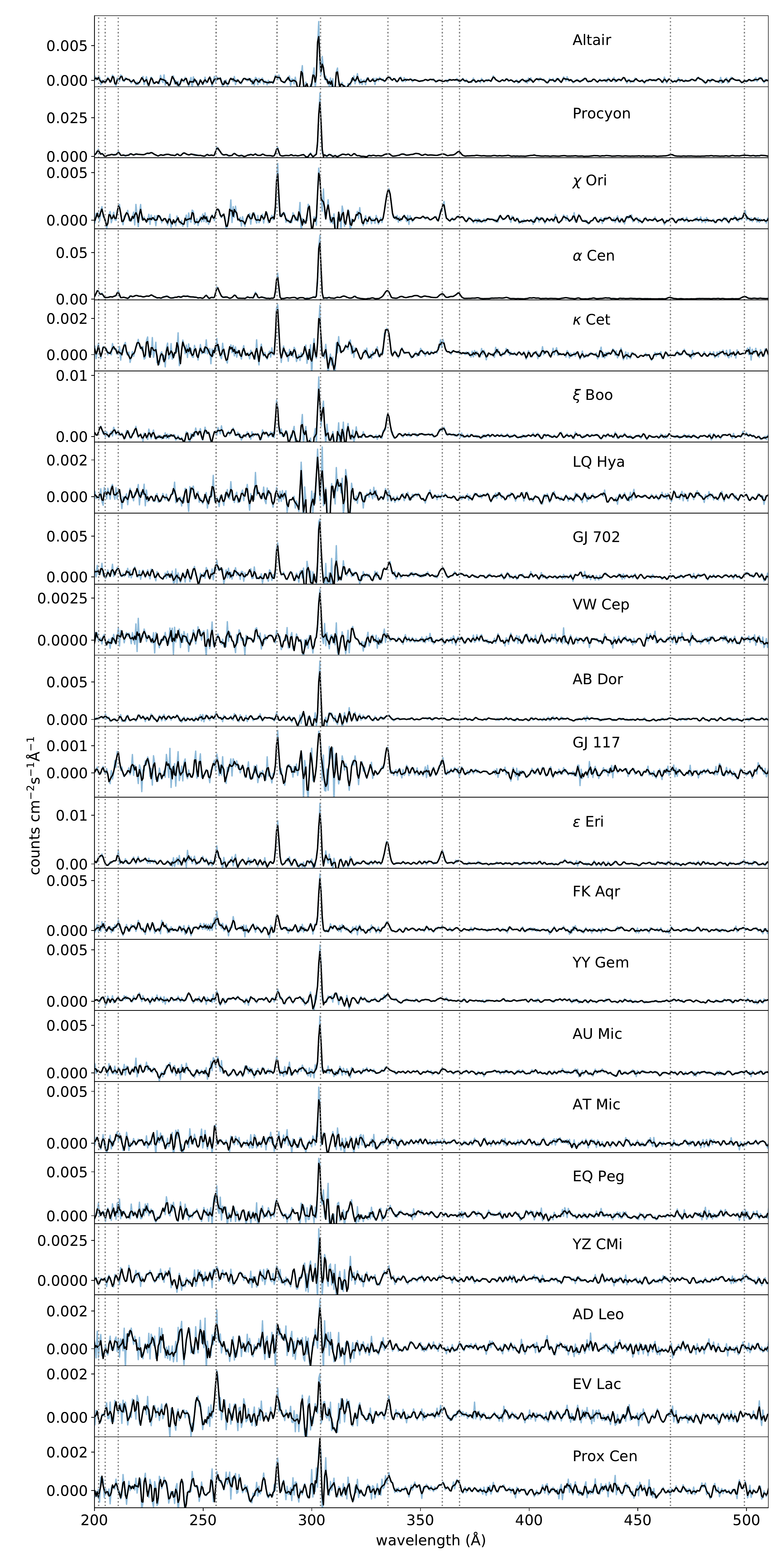}
\caption{EUVE spectra of the sample stars in the 200-504\,\AA\ wavelength range. The light blue lines show the unbinned spectra, the black lines show the spectra smoothed by a Savitzky-Golay filter with a window length of 5, and the vertical dotted lines show the wavelengths of the strong stellar emission lines from Table~\ref{tab:lines}. }
\label{fig:atlas}
\end{figure*}

Fig.~\ref{fig:atlas} shows the EUVE spectra of the sample stars.

\section{Source tables}

Information on the properties of the sample stars and their narrow-band EUVE luminosity is given in Table~\ref{tab:sample} and their X-ray spectral properties are given in Table~\ref{tab:specfits}.

\begin{table*}
\caption{Properties of the observed stars. Distances are from \citet{Bailer-Jones2018} by default, and from Hipparcos when marked with (H). }
\label{tab:sample}
\begin{tabular}{l l l l l l l l }
\hline \hline
Name & Other name & Spectral type & Distance (pc) & Observation ID & Detector used for  & X-ray & $L_{\mathrm{200-504} \mbox{\tiny{\normalfont\AA}}  }$\\
 &  &  &  &  & X-ray spectra &  pile-up? & (\lumi )\\

\hline
Altair         & HD 187642   & A7 V       & 5.13 (H)    & XMM 0502360101  & PN thick & no & $3.3\times 10^{27}$\\
Procyon        & HD 61421    & F5 IV-V    & 3.51 (H)    & XMM 0123940101  & PN thick & no & $1.2\times 10^{28}$ \\
$\chi$ Ori     & HD 39587    & G0 V       & 8.84    & XMM 0111500101  & MOS2 thick & no & $2.5\times 10^{28}$ \\
$\alpha$ Cen AB& HD 128620   & G2 V       & 1.35 (H)    & XMM 0045340901  & PN thick & no & $4.2\times 10^{27}$ \\
$\kappa$ Cet   & HD 20630    & G5 V       & 9.14    & XMM 0111410101  & MOS1 thick & no & $1.0\times 10^{28}$ \\
$\xi$ Boo      & HD 131156   & G8 V       & 6.73    & XMM 0056030101  & MOS2 medium & no & $1.7\times 10^{28}$ \\
LQ Hya         & HD 82558    & K0V        & 18.28    & XMM 0148880101  & PN medium & no & $1.6\times 10^{28}$ \\
GJ 702AB       & HD 165341   & K0V        & 5.08    & XMM 0044741301  & PN thick & no & $7.0\times 10^{27}$ \\
VW Cep         & HD 197433   & K0V        & 26.60    & XMM 0100650301  & PN thick & no & $5.1\times 10^{28}$ \\
AB Dor         & HD 36705    & K1 IIIp    & 15.30    & XMM 0412580801  & PN thick & yes & $2.4\times 10^{28}$ \\
GJ 117         & HD 17925    & K2V        & 10.355    & XMM 0203060501  & PN thick & no & $9.5\times 10^{27}$ \\
$\epsilon$ Eri & GJ 144      & K2V        & 3.203    & XMM 0820070201  & PN thick & no & $5.4\times 10^{27}$ \\
FK Aqr         & GJ 867AB    & M0 Vpe     & 8.897    & XMM 0151450101  & PN medium & yes & $1.4\times 10^{28}$ \\ 
YY Gem         & GJ 278C     & M1 Ve      & 15.09    & XMM 0112880801  & PN thick & yes & $2.9\times 10^{28}$ \\ 
AU Mic         & GJ 803      & M2 Ve      & 9.722    & XMM 0111420101  & PN medium & yes & $1.5\times 10^{28}$ \\ 
AT Mic AB      & GJ 799AB    & M4 Ve      & 9.85    & XMM 0111510101  & PN medium & yes & $1.2\times 10^{28}$ \\ 
EQ Peg         & GJ 896AB    & M4 V       & 6.260    & XMM 0112880301  & PN thick & yes & $9.1\times 10^{27}$ \\
YZ CMi         & GJ 285      & M4.5 Ve    & 5.986    & XMM 0111460101  & PN medium & no & $3.1\times 10^{27}$ \\
AD Leo         & GJ 338      & M4.5 Ve    & 4.965    & XMM 0111440101  & PN medium & yes & $2.4\times 10^{27}$ \\
EV Lac         & GJ 873      & M4.5 Ve    & 5.049    & XMM 0111450101  & PN medium & yes & $3.2\times 10^{27}$\\
Prox Cen       & GJ 551      & M5 Ve      & 1.301    & XMM 0551120201  & PN medium & no & $1.5\times 10^{26}$ \\

\hline
\end{tabular}
\end{table*}

\begin{landscape}

\begin{table}
\caption{Fitted X-ray spectral parameters}
\label{tab:specfits}
\begin{small}
\begin{tabular}{lrrrrrrrrrrrr}
\hline\hline
Name & EM$_\mathrm{1\,MK}$\footnotemark & EM$_\mathrm{2\,MK}$ & EM$_\mathrm{4\,MK}$ & EM$_\mathrm{8\,MK}$ & EM$_\mathrm{16\,MK}$ & EM$_\mathrm{32\,MK}$ & T$_{\mathrm{cor,\, mean}}$ & [Fe/H]$_{\mathrm{cor}}$ & [O/H]$_{\mathrm{cor}}$ & [Ne/H]$_{\mathrm{cor}}$ & $F_\mathrm{X,\,0.2-5\,keV}$ & $L_\mathrm{X,\,0.2-5\,keV}$ \\
 & & & & & & & (10$^6$ K) & & & & (10$^{-12}$\,\flux) & (10$^{26}$\lumi) \\

\hline

Altair & -- & 2.6$\pm$0.1 & 0.4$\pm$0.1 & -- & -- & -- & 2.3 & 1.4$\pm$0.25 & 0.6$\pm$0.03 & 1.0$\pm$0.13 & 0.6 & 1.2 \\
Procyon & -- & 37.2$\pm$0.3 & -- & 0.2$\pm$0.1 & -- & -- & 2.1 & 1.8$\pm$0.09 & 0.4$\pm$0.01 & 0.5$\pm$0.04 & 6.6 & 14.2 \\
$\chi$ Ori & -- & 2.5$\pm$1.0 & 31.7$\pm$1.6 & 16.1$\pm$0.8 & 1.0$\pm$0.7 & 1.0$\pm$0.5 & 6.0 & 0.9$\pm$0.03 & 0.4$\pm$0.03 & 0.5$\pm$0.05 & 9.6 & 20.8 \\
$\alpha$ Cen & -- & 56.4$\pm$1.5 & 12.9$\pm$2.0 & 2.8$\pm$0.5 & -- & -- & 2.6 & 1.4$\pm$0.13 & 0.4$\pm$0.02 & 0.6$\pm$0.10 & 13.1 & 28.5 \\
$\kappa$ Cet & -- & 3.0$\pm$0.9 & 20.8$\pm$1.5 & 11.6$\pm$0.8 & 1.7$\pm$0.4 & -- & 5.7 & 1.1$\pm$0.05 & 0.6$\pm$0.04 & 1.4$\pm$0.08 & 8.6 & 18.6 \\
$\xi$ Boo & -- & 14.7$\pm$0.7 & 39.2$\pm$1.3 & 28.5$\pm$0.9 & 7.7$\pm$0.8 & 0.9$\pm$0.5 & 6.3 & 0.7$\pm$0.02 & 0.5$\pm$0.01 & 1.3$\pm$0.03 & 17.1 & 37.1 \\
LQ Hya & -- & 7.6$\pm$0.4 & 40.1$\pm$1.1 & 28.9$\pm$1.0 & 12.2$\pm$1.0 & 14.4$\pm$0.7 & 10.5 & 0.3$\pm$0.01 & 0.6$\pm$0.01 & 1.7$\pm$0.03 & 16.1 & 35.0 \\
GJ 702 & -- & 10.8$\pm$0.9 & 15.3$\pm$1.7 & 9.1$\pm$1.1 & 0.7$\pm$0.6 & -- & 4.7 & 0.7$\pm$0.05 & 0.5$\pm$0.03 & 1.0$\pm$0.09 & 6.5 & 14.1 \\
VW Cep & -- & 1.4$\pm$0.7 & 13.3$\pm$1.8 & 21.8$\pm$1.7 & -- & 5.8$\pm$0.9 & 10.0 & 0.4$\pm$0.02 & 0.8$\pm$0.06 & 2.0$\pm$0.12 & 7.5 & 16.2 \\
AB Dor & -- & 15.2$\pm$3.9 & 103.3$\pm$8.8 & 71.2$\pm$7.1 & 164.7$\pm$4.0 & -- & 10.5 & 0.4$\pm$0.02 & 0.6$\pm$0.03 & 1.4$\pm$0.06 & 53.3 & 115.8 \\
GJ 117 & -- & 4.1$\pm$0.4 & 20.6$\pm$0.8 & 9.5$\pm$0.5 & -- & 1.2$\pm$0.3 & 5.9 & 0.7$\pm$0.02 & 0.7$\pm$0.02 & 1.3$\pm$0.04 & 6.7 & 14.6 \\
$\epsilon$ Eri & -- & 23.0$\pm$1.1 & 61.7$\pm$2.2 & 24.6$\pm$1.3 & -- & 1.0$\pm$0.6 & 4.8 & 0.7$\pm$0.02 & 0.6$\pm$0.02 & 1.2$\pm$0.03 & 20.9 & 45.3 \\
FK Aqr & 3.8$\pm$0.4 & 15.7$\pm$0.5 & 46.6$\pm$1.2 & 21.7$\pm$1.1 & 16.6$\pm$0.8 & -- & 6.4 & 0.4$\pm$0.01 & 0.7$\pm$0.01 & 1.5$\pm$0.03 & 17.2 & 37.4 \\
YY Gem & -- & 8.1$\pm$1.5 & 40.3$\pm$2.5 & 17.7$\pm$2.0 & 15.9$\pm$1.9 & 17.2$\pm$1.3 & 11.4 & 0.4$\pm$0.02 & 0.7$\pm$0.04 & 1.8$\pm$0.07 & 16.1 & 35.0 \\
AU Mic & -- & 14.0$\pm$1.3 & 64.4$\pm$2.0 & 26.8$\pm$1.6 & 32.0$\pm$0.9 & -- & 7.5 & 0.4$\pm$0.01 & 0.7$\pm$0.02 & 1.6$\pm$0.03 & 22.1 & 48.0 \\
AT Mic & -- & 10.7$\pm$1.2 & 61.6$\pm$2.5 & 19.5$\pm$2.1 & 46.6$\pm$1.2 & -- & 8.6 & 0.3$\pm$0.01 & 0.7$\pm$0.02 & 1.4$\pm$0.03 & 21.0 & 45.6 \\
EQ Peg & -- & 16.5$\pm$1.2 & 29.1$\pm$2.7 & 28.4$\pm$2.3 & 17.1$\pm$1.2 & -- & 7.3 & 0.5$\pm$0.02 & 0.8$\pm$0.04 & 1.6$\pm$0.07 & 16.4 & 35.6 \\
YZ CMi & -- & 9.9$\pm$0.8 & 20.9$\pm$1.0 & 11.5$\pm$0.9 & 5.3$\pm$0.7 & 3.0$\pm$0.4 & 7.5 & 0.5$\pm$0.02 & 0.8$\pm$0.03 & 1.4$\pm$0.05 & 8.7 & 18.8 \\
AD Leo & -- & 16.5$\pm$1.2 & 43.5$\pm$1.6 & 25.2$\pm$1.5 & 9.9$\pm$1.4 & 5.7$\pm$0.9 & 7.6 & 0.4$\pm$0.01 & 0.7$\pm$0.02 & 1.7$\pm$0.04 & 17.1 & 37.1 \\
EV Lac & -- & 14.9$\pm$1.3 & 41.4$\pm$1.8 & 25.9$\pm$1.5 & 19.1$\pm$0.8 & -- & 7.1 & 0.4$\pm$0.01 & 0.7$\pm$0.02 & 1.3$\pm$0.04 & 16.6 & 36.2 \\
Prox Cen & -- & 6.1$\pm$0.2 & 9.1$\pm$0.5 & 5.0$\pm$0.4 & 0.8$\pm$0.4 & -- & 5.0 & 0.4$\pm$0.02 & 0.7$\pm$0.02 & 1.2$\pm$0.05 & 3.5 & 7.7 \\

\hline
\end{tabular}
\end{small}
\end{table}

\footnotetext{Emission measures are given here as the Xspec "norm" parameter of the fitted "apec" model. They can be transformed to physical quantities in units of cm$^{-3}$ by multiplying them with $4\pi d^2 \times 10^{14}$, with $d$ being the distance to the star in cm.}

\end{landscape}



\newpage
\bsp	
\label{lastpage}
\end{document}